\documentclass[fleqn,usenatbib]{mnras}

\usepackage[T1]{fontenc}
\usepackage{ae,aecompl}

\usepackage{textcomp}
\usepackage[dvips]{graphicx}
\usepackage{graphicx}
\usepackage{amsmath,amssymb}
\usepackage{epsf}
\usepackage{dcolumn}
\usepackage{multirow}
\usepackage{epsf}
\usepackage{bm}
\usepackage[utf8]{inputenc}  
\usepackage{txfonts}  
\usepackage{url}
\usepackage[dvipsnames]{xcolor}
\usepackage{color}

\newcommand{\be}{\begin{equation}}
\newcommand{\ee}{\end{equation}}
\newcommand{\rockstar}{\textsc{Rockstar}}

\begin{document}
\title[The Galaxy Halo Connection in Modified Gravity Cosmologies]{The Galaxy Halo Connection in Modified Gravity Cosmologies:
Environment Dependence of Galaxy Luminosity function}

\author[N. C. Devi et. al]{N. Chandrachani Devi$^{1,2}\thanks{E-mail: chandrachani@gmail.com}$,
Aldo Rodr\'iguez-Puebla$^1$,
O. Valenzuela$^1$, 
Vladimir Avila-Reese$^1$,
\newauthor{César Hernández-Aguayo$^3$ 
and
Baojiu Li$^3$}
\\
$^1$ Instituto de Astronom\'ia, Universidad Nacional Aut\'onoma de M\'exico, A. P. 70-264, 04510, M\'exico, D.F., M\'exico\\
$^{2}$Instituto de F\'isica, Universidad Nacional Autonoma de M\'exico, Circuito de la Investigación Científica Ciudad Universitaria, 04510, M\'exico, D.F., M\'exico\\
$^3$Institute for Computational Cosmology, Department of Physics, Durham University, South Road, Durham DH1 3LE, UK.\\
} 
 
\date{\today}

\maketitle

\begin{abstract}
We investigate the dependence of the galaxy-halo connection and galaxy density field in modified gravity models using the $N-$body simulations for $f(R)$ and nDGP models at $z=0$. Because of the screening mechanisms employed by these models, chameleon and Vainshtein, halos are clustered differently in the non-linear regime of structure formation. We quantify their deviations in the galaxy density field from the standard $\Lambda$CDM model under different environments. We populate galaxies in halos via the (Sub)Halo Abundance Matching. Our main
results are: 1) The galaxy-halo connection {\it strongly} depends on the gravity model;  a maximum variation of $\sim40\%$ is observed between Halo
Occupational Distribution (HOD) parameters; 2) $f(R)$ gravity models predict an excess of galaxies in low density environments of $\sim10\%$ but predict a deficit of $\sim10\%$ at high density environments for $|f_{R0}| = 10^{-4}$ and $10^{-6}$ while $|f_{R0}| = 10^{-5}$ predicts more high density structures; nDGP models are consistent with $\Lambda$CDM; 3) Different gravity models predict different dependences of the galaxy luminosity function (GLF) with the
environment, especially in void-like regions we find differences around $\sim10\%$ for the
$f(R)$ models while nDPG models remain closer to $\Lambda$CDM for low-luminosity galaxies
but there is a deficit of $\sim11\%$ for high-luminosity galaxies in all environments.  We conclude that the dependence of the GLF with environment might provide a test to distinguish between gravity models and their screening mechanisms from the $\Lambda$CDM. We provide HOD parameters for the gravity models analyzed in this paper.

\end{abstract}
 
\begin{keywords}
Cosmology: Large Scale structure of Universe, Dark Energy, Dark Matter. Galaxies: haloes,  Luminosity function.

\end{keywords}

\section{Introduction}

The standard model of cosmology assumes that our universe is homogeneous and isotropic and
that our current knowledge of gravity is well described by general relativity(GR). With the recent discovery of the late-time accelerated expansion \citep{Riess:1998cb,Perlmutter:1998np},
the most popular approach to describe the dynamics of our universe within
the general relativity framework is by introducing the hypothesis of a dark energy component with a negative pressure permeating all over the space. Among the various explanations proposed, there are mainly two candidates for dark energy well studied in literature. One are quintessence models in which the dark energy is governed by a dynamical scalar field \citep{Ratra+1988}. The other candidate is the cosmological constant $\Lambda$, incarned as a vacuum energy that remains after the inflationary epoch \citep[for a review see][]{Carroll2001}. Yet the $\Lambda$ Cold Dark Matter model, $\Lambda$CDM, remains as the most popular and widely accepted cosmological gravity model. 

The $\Lambda$CDM model is highly successful in explaining a number of cosmological probes such as the temperature anisotropies and polarization in the cosmic microwave background radiation (CMBR), the baryon acoustic oscillations (BAO) imprinted in the galaxy spatial distributions at large scales, and the accelerated expansion inferred mainly from observations of
high-redshift Type Ia supernovae \citep{Ade+2016,Aghanim+2018}. The $\Lambda$CDM model provides also a successful background to explain a number of astronomical observations such as the galaxy clustering both for local and high redshift galaxies \citep[e.g.,][see for a review, \citealp{FrenkWhite2012}]{Conroy+2006,Reddick+2012}, the galaxy cluster mass function \citep[e.g.,][]{Vikhlinin+2009}, the so-called Star-Forming Main Sequence of galaxies at different redshifts \citep[e.g.,][]{Behroozi2013b,Rodriguez+2016b}, and a number of galaxy demographics and relationships that have been well studied based on analytic and semi-analytic models \citep[e.g.,][]{Mo+1998,Firmani+2000,Baugh+2006} and large cosmological hydrodynamics simulations \citep{Vogelsberger+2014,Schaye+2015} of galaxy formation \citep[for a review, see][and more references therein]{Somerville+2015}. 

Despite of the success of the 
$\Lambda$CDM cosmology, so far, we do not have a solution to the problems that raises the explanation of the origin of $\Lambda$ \citep{Weinberg1989,Sahni:1999gb}, nor
GR has been tested on cosmological scales, though there are some hints
of being valid at galactic scales \citep{Collett+2018}. In addition, the recently reported
$\sim3.5\sigma$ discrepancy between the Hubble constant, $H_{0}$, locally determined \citep{Beaton:2016nsw,Freedman:2017yms,Riess:2018byc} 
and the one estimated from the CMBR anisotropies \citep{Ade+2016,Aghanim+2018},
has led to a tension within the standard cosmology \citep{Riess+2016}; but see \citet{Shanks+2018} for
a possible solution for the above tension and \citet{Riess+2018f} for
a counterargument to that paper. The above has been interpreted as the necessity 
to consider a revision of our knowledge on gravity over cosmological scales or
perhaps new physics. 

\subsection{Modified Gravity models and their screening mechanisms}

Recently, alternative modified gravity (MG) models have received a lot of attention 
as they offer interesting and possible alternative explanations to the cosmic 
acceleration of the Universe, without invoking dark energy  but by
naturally modifying GR on cosmological scales. 
Nonetheless, even when these models are modification to GR on the large scales,
they still need to satisfy the tight constraints on deviations from GR from the Solar System \citep{Will:2014kxa}.
Screening mechanisms have helped to overcome the potential inconsistency; some of them were developed
with the aim to suppress efficiently the fields that mediate the MG force 
at small scales while enhancing the effect of gravity over cosmological 
scales (Vainshtein models) or at scales smaller than the scalar field Compton wavelength \citep{Khoury:2010xi,Joyce:2014kja}. 

In this work, we will explore two types of MG models: the normal \citet{Dvali2000} (DGP) model and 
the chameleon $f(R)$ model \citep{HuSawicky2007}. The former assumes
that the standard model of particles live in a 4D spacetime brane embedded within a 
higher dimensional 5D space in which only gravity propagates. The 
latter introduces an arbitrary function of the Ricci scalar that generalizes
the Einstein-Hilbert action; here, we consider the functional form proposed by 
\citet{HuSawicky2007}. While there are several
other classes of MG models, we explore here only the above two models
as they are the most popular and widely studied proposals so far
\citep[for a review see][]{Koyama2016}. 

As for the screening mechanisms, there are two types well studied, the 
Vainshtein and the chameleon ones. In this work, the former is applied to the DGP model
and the latter to the $f(R)$ model. In both
models, the screening mechanism is governed by an
extra degree of freedom introduced, most of the time, as a scalar field
that follows a non-linear equation strongly coupled to the density field. 
As a result, one expects that the clustering of the dark matter particles
is different between these two screening mechanisms;
the Vainshtein mechanism depends on the locations of the particles in the cosmic web while in chameleon mechanism is
roughly independent \citep{Falck:2015rsa}. This leads to the speculation
that screening mechanisms might affect the structure and spatial distribution of
dark matter halos. 
Indeed, recent studies have shown that screening mechanisms affect 
halos differently. Based on high-resolution N-body simulations
\citet{Falck:2015rsa} showed that the Vainshtein screening mechanism
is independent of halo mass and is very efficient inside the virial radius
but decreases at larger radii. Instead, the chameleon mechanism
depends on halo mass and the strength of $f_{R0}$, as well as on halo screening 
profiles; for similar conclusions see 
\citep{Zhao:2011cu,Shi:2017pyd,Winther:2011qb}.
Thus, it is of interest to see whether such effects can be
detected observationally in the correlations and statistical distributions of galaxies and cluster of galaxies. 

\subsection{Exploring the effects of MG models on the galaxy distributions}

Based on the discussion above, 
we propose to study here the galaxy density field under the
$f(R)$ and DGP MG models and their respective screening mechanisms. 
Galaxies are biased tracers of dark matter halos, and dark matter halos are biased tracers
of the dark matter particles  distribution. The latter is due to the highly non-linear evolution
of the mass density perturbation field, and the former is consequence of the complex
gastro-physical processes that govern galaxy formation and evolution \citep{Somerville+2015}.
While galaxy formation within the evolving dark matter halos remains as one of the most challenging 
problems in modern astronomy, the statistical matching between observed galaxies and simulated dark matter halos allows 
for a direct connection of galaxies to halos. This connection has been
very well constrained in the past not only for local
galaxies but up to very high redshifts  
(see e.g., \citealt{Rodriguez-Puebla+2017,Behroozi+2018};
for a recent review, see \citealt{Wechsler2018}). 
Thus, the results from N-body simulations of structure formation under MG cosmologies can be connected statistically to galaxies, in such a way that the measured galaxy correlations and spatial distributions in the simulations can be compared with observations. 

Using galaxies to constrain MG models is not a new idea \citep[][see for a review of various
astrophysical test using galaxies,]{Koyama2016}. Indeed,
present and future spectroscopic and imaging galaxy surveys, 
such as the extended Baryon Oscillation Spectroscopic 
Survey (eBOSS)\footnote{https://www.sdss.org/surveys/eboss/}, the dark energy survey(DES) \footnote{https://www.darkenergysurvey.org/es/}, the Dark Energy Spectroscopic Instrument (DESI) \footnote{https://www.desi.lbl.gov/} \citep{Aghamousa:2016zmz}, and the European Space 
Agency’s-Euclid\footnote{https://www.euclid-ec.org/} will not only allow to test the gravity theory on the largest scales of our Universe but also will allow to constrain a wide range of cosmological scenarios. 

The main goal of this work is to quantify the effects of the different gravity models discussed above on the statistics of the dark matter halos, and its implication on the distributions of their host galaxies.

In particular, we will do so by studying their density field via the galaxy luminosity function  (GLF) at $z\approx 0$, when the effects of environments are more relevant.
This way, we will evaluate whether observational tests, as the variation of the GLF with environment, could be useful for discriminating the studied MG models.

In this paper, we generate galaxy mock catalogs using N-body simulations via the (sub)halo abundance matching (SHAM) technique. 
We will focus mainly on the results based on the GLF in the $r-$band and its dependence on large-scale environment.  The viability of our mock catalogs is tested
by showing that all the models produce two-point correlation function that are in agreement
with the observations from the SDSS DR7 \citep{Zehavi2011}.
We also provide HOD parameters for all our galaxy mocks based on different MG models. 

This paper is organized as follows. In Section \ref{sec:Models_mod_grav}, 
we describe the MG models we employed: $f(R)$ and DGP models. In Section \ref{sec:N-body} 
we describe the suite of simulations we use for the MG models, previously
described in \citet{Li2012} and \citet{Li2013}. 
Section \ref{sec:galaxy-halo_connection} describes the SHAM approach we use for the galaxy-halo
connection and we show that all our galaxy mocks produce realistic two point correlation function. 
In Section \ref{sec:results_section} we present
our results on the predicted galaxy density field
and the dependence of the GLF
with environment. In Section \ref{secc:discusion} we
discuss our results. 
Finally, Section \ref{sec:summary_and_discussion} 
we present a summary with our main results and a 
discussion.

The cosmological parameters used for this paper are:
$\Omega_{\rm m} = 0.281$, $\Omega_{\Lambda} = 1 - \Omega_{\rm m}$, $n_s = 0.971$ and $\sigma_8 = 0.820$. 

\section{Modified Gravity models}
\label{sec:Models_mod_grav}

In this Section, we briefly describe the two models of MG that we will use in this paper.

\subsection{$f(R)$ gravity model}

Firstly, we consider the $f(R)$ theories of gravity, where the Ricci scalar $R$ in the Einstein-Hilbert action is generalized by a functional form  $f(R)$: 
\begin{equation}
S=\frac{1}{2\kappa}\int{d^{4}x\sqrt{-g}(R+f(R))}+S_{m}(g_{\mu\nu},\psi_{m})\,,
\label{eq:action}
\end{equation}
where $\kappa=8\pi G$, $G$ is the Newtonian gravitational constant, $g$ is the 
determinant of the metric tensor $g_{\mu\nu}$, $f(R)$ is an arbitrary function of
the Ricci scalar $R$ and $S_{m}$ is the matter action that depends on $g_{\mu\nu}$
and matter fields $\psi_{m}$.

By varying the action with respect to the metric $g_{\mu\nu}$, we obtain the modified Einstein equations \citep{DeFelice:2010aj}

\begin{equation} 
\label{eq:field metric}
f_{R}(R)R_{\mu\nu}-\frac{1}{2}f(R)g_{\mu\nu}-\nabla_{\mu}\nabla_{\nu}f_{R}(R)+g_{\mu\nu}\Box f_{R}(R)=kT_{\mu\nu},
\end{equation}
where $f_{R}(R)=\mathrm{d}f(R)/\mathrm{d}R$ represents the extra degree of freedom,
i.e., the scalar field and often known as the scalaron field.  The d'Alambertian operator
is denoted with $\Box=\nabla_{\alpha}\nabla^{\alpha}$ and $\nabla_\alpha$ is the usual
covariant derivative associated with respect to the affine connections of the metric 
while $T_{\mu\nu}$ is the energy-momentum tensor of the matter fields.

Thus, under the scalar field representation of $f(R)$, the equation of motion that determines the dynamics of scalar field, $f_{R}$, is given by the trace of Eq.~\eqref{eq:action}
\begin{equation}
\Box f_{R} = \frac{1}{3}\left[R - f_{R} R + 2f(R) - 8\pi G \rho_{m}\right]=\frac{\partial V_{\rm eff}}{\partial f_{R}}.
\label{eq:field_Eq}
\end{equation}
Here $\rho_{m}$ is the non-relativistic matter (including dark matter
and baryons) density of the Universe. 
The structure formation in the non-linear regime is well studied by assuming the quasi-static and weak-field approximations \citep{Bose:2014zba}. Under such limits 
Eq.~\eqref{eq:field metric} reduces to the Poisson equation
\begin{equation}
\nabla^2 \Phi = 4\pi G a^2 \delta\rho_{m} - \frac{1}{2}\nabla^2f_{R},
\label{eq:poisson}
\end{equation} 
and Eq.~\eqref{eq:field_Eq} is
\begin{equation}
\nabla^2 f_{R} = \frac{a^2}{3}\left(\delta R(f_{R}) - 8 \pi G \delta \rho_{m}\right),
\end{equation}
where $\Phi$ represents the gravitational potential at some particular position in
the space and corresponding to the density fluctuation $\delta \rho_{m} = \rho_{m}- \bar{\rho}_{m}$ with curvature perturbation $\delta R = R - \bar{R}$. Here $\bar{\rho}_{m}$ and $\bar{R}$ are the background matter density and the curvature of the Universe, respectively. This system of equations determines the effect of MG on the structure formation. While in case of $\Lambda$CDM model, 
the Poisson equation has form $\nabla^2 \Phi = 4\pi G a^2 \delta\rho_{m}$. Therefore, the term $\nabla^2 f_{R}$ in eq.(\ref{eq:poisson}) drives the  modified gravity effect in comparison to $\Lambda$CDM model.

Different forms of $f(R)$ gravity models have been proposed in the literature \citep[see e.g.,][]{Cognola2008,Linder2009, HuSawicky2007, Capozziello2002, Nojiri2003,Dolgov+2003,Faraoni+2006,Chiba+2003}. 

For this work, we consider the most widely studied functional form of $f(R)$, proposed by \citet{HuSawicky2007} which satisfies both cosmological and solar-system tests \citep{Martinelli2009}
\begin{equation}
f(R)=-m^{2}\frac{ c_{1}\left(-\frac{R}{m^{2}}\right)^{n} }{c_{2}\left(-\frac{R}{m^{2}}\right)^{n}+1}.
\label{HSmodel}
\end{equation}
Here $n$, $c_{1}$ and $c_{2} $ are dimensionless model parameters and $m^{2} = H^{2}_{0}\Omega_{m0}$ with $H_{0}$ and $\Omega_{m0}$ are respectively the present day values of the Hubble constant and the matter density parameters of the Universe. If we set $c_1/c_2 = \Omega_{\Lambda 0}/\Omega_{m0}$, with $\Omega_{\Lambda0} \equiv 1 - \Omega_{m0}$, the model is able to mimic the expansion history of $\Lambda$CDM. From Eq.~\eqref{HSmodel} the functional form of the scalaron field, $f_R$, is given by
\begin{equation}
f_{R} = - \frac{c_{1}}{c^{2}_{2}}\frac{n(-R/m^2)^{n-1}}{\left[(-R/m^2)^n+ 1\right]^2}\,, \label{eq:f_R}
\end{equation}
where
\begin{equation}\label{eq:c1c2}
\frac{c_{1}}{c_{2}^{2}} = -\frac{1}{n}\left[3\left(1+4\frac{\Omega_{\Lambda0}} {\Omega_{m0}}\right)\right]^{(n+1)} f_{R0}\,,
\end{equation}
with $f_{R0}$ is the present value of the scalaron field.

We notice that from Eq.~\eqref{eq:c1c2}, $n$ and $f_{R0}$ are the remaining free parameters of the model. Hence, in this work we adopt the values $n=1$ and $|f_{R0}| = 10^{-6}$, $10^{-5}$ and $10^{-4}$ (hereafter referred as F6, F5 and F4, respectively) which correspond to a weak, medium and strong deviation with respect to GR, i.e., to $\Lambda$CDM model.  

The aim of this work is to study the effect of MG models under different density environments. Thus, it is important to understand the screening mechanism implemented in such models that could have a direct impact in the halo (and thus galaxy) density field. Recall
that the screening mechanisms are applied in MG models to suppress the enhancement of the fifth force in order to pass high precision tests of gravity in the high dense regions like the Solar System \citep[see e.g.,][]{Vainshtein:1972sx,Khoury:2003rn}.
To suppress the effects of the fifth force in high density regions, the Hu-Sawicki $f(R)$ gravity model employs the chameleon mechanism \citep{Khoury:2003rn}, where the scalaron field which mediates the fifth force has a non-zero mass, $m_{f_R}^2 = \partial^2 V_{\rm eff}/\partial^2 f_{R}$, depending on the non-linear terms in the equation of motion, see
Eq. (\ref{eq:field_Eq}). The corresponding fifth force is of Yukawa-type, decaying exponentially with mass as $\propto\exp(-m_{f_R}r)$ where $r$ is the separation between two test masses. Under the high density environments, this scalaron field is massive and because of it, the fifth force is suppressed sufficiently and  allow to recover GR successfully.
Depending on the requirement of screening different density environments of our universe, various constrains on the $f_{R0}$ values are being studied in the literature \citep[for recent reviews see]{Jain+2012,Vikram+2013,Lombriser:2014dua,Burrage:2016bwy}.

{\color{black} Recent studies based on 
Milky-Way type galaxies, being need to be screened in order to satisfy the Solar System test, imposes the constraint on the present day value
for the scalaron field to be within the range of 
$|f_{R0}|<10^{-4} - 10^{-6}$ \citep{HuSawicky2007},
while for dwarf galaxies, it has been reported the values as low as $|f_{R0}| \lesssim 10^{-7}$  
\citep{Lombriser:2014dua}. On the other hand, the
abundance of galaxy clusters, based on X-ray data, in
combination with the CMBR, based on the Planck measurement,
and SNIa and BAOs, \cite{Cataneo:2014kaa} 
found an
upper value of $|f_{R0}| < 1.6\times 10^{-5}$ at the 95.4 per cent confidence level. 
}

\subsection{nDGP model}

In the Dvali–Gabadadze–Porrati (DGP) braneworld cosmological model \citep{Dvali2000}, the standard 4-dimensional spacetime universe (or brane) is embedded in a 5-dimensional bulk space-time with an infinite extra dimension; the standard matter particles are confined only on the brane surface. In this model, the graviton field is freely propagated into the extra dimension \citep{Koyama:2007ih}.
Thus, the modifications of Einstein’s general relativity are quantified in the action:

\begin{equation}
S = \int_{\rm brane} d^4x \sqrt{-g} \frac{R}{2\kappa} +\int_{\rm bulk}d^5x \sqrt{-g^{(5)}} \frac{R^{(5)}}{2\kappa^{(5)}} + S_{m}(g_{\mu\nu},\psi_{m}),
\end{equation}
where $\kappa^{(5)} = 8 \pi G^{(5)}$ with $G^{(5)}$ being the 5-D gravitational constant and $R^{(5)}$ is the Ricci scalar in 5 dimensions. The gravitational constants in the brane $(G)$ and in the bulk $(G^{(5)})$ are related through the cross-over scale, $r_c$,
\begin{equation}
r_c = \frac{1}{2} \frac{G^{(5)}}{G}\,,
\end{equation} 
where below this scale the strength of four-dimensional gravity becomes dominant and vise versa.

Here, we consider the {\it normal branch} of the DGP model, nDGP, where a dark energy component is included to the matter part of action in order to have an accelerated expansion of the Universe \citep{Schmidt:2009sv}. Hence the expansion rate under such scenario is given by
\begin{equation}\label{eq:H_nDGP}
H(a) = H_0\left( \sqrt{\Omega_{\rm m}a^{-3} + \Omega_{\rm DE}(a) + \Omega_{\rm rc}} - \sqrt{\Omega_{\rm rc}}\right)\,,
\end{equation}
with $\Omega_{\rm DE}$ is the contribution of the dark energy component and $\Omega_{\rm rc} = 1/(4H^2_0r^2_c)$ is a dimensionless parameter related to the cross-over scale. From Eq.~\eqref{eq:H_nDGP} we notice that in the limit $\Omega_{\rm rc} \to 0$ or $H_0r_c \to \infty$ we recover the expansion history of the $\Lambda$CDM model. Here, we consider two variants of the nDGP model with $H_0r_c = 5$, N5, and $H_0r_c=1$, N1, which represent models with a weak and medium deviation with respect to GR, respectively.

Similar to $f(R)$ gravity models, the appearance of a new scalar field, $\varphi$, associated with the bending modes of the 4D brane mediates the fifth force. In this model, the modified Poisson equation and the equation of motion for the scalar field $\varphi$ under the quasi-static approximation are given by \citep{Koyama:2007ih}
\begin{equation}
\nabla^2 \Phi = 4\pi G a^2 \delta \rho_{\rm m} + \frac{1}{2}\nabla^2\varphi,
\label{eq:poisson_dgp}
\end{equation}
with
\begin{equation}\label{eq:phi_dgp}
\nabla^2 \varphi + \frac{r_c^2}{3\beta\,a^2} \left[ (\nabla^2\varphi)^2 - (\nabla_i\nabla_j\varphi)^2 \right] = \frac{8\pi\,G\,a^2}{3\beta} \delta\rho_{\rm m},
\end{equation}
and
\begin{equation}
\beta = 1 + 2 H\, r_c \left( 1 + \frac{\dot H}{3 H^2} \right).
\end{equation}
Here, the term $\nabla^2\varphi$ in the Poisson equation Eq.~\eqref{eq:poisson_dgp} governs the MG effect.
One can understand the screening mechanism employed in this model by simply considering the spherically symmetric geometry in which the field equation, Eq.~(\ref{eq:phi_dgp}), has an analytic solution given by \citep{Koyama:2007ih}:
\begin{equation}
\varphi_{,r} = \frac{4}{3\beta}\left(\frac{r}{r_{V}}\right)^3\left[-1 +\sqrt{1+\left(\frac{r_{V}}{r}\right)^3} \right]\frac{GM(r)}{r^2}.
\end{equation}
Here $r_V$ is a distance scale called 
Vainshtein radius defined as: 
\begin{equation}\label{eq:r_V}
r_V(r) = \left(\frac{16r^2_c GM(r)}{9\beta^2}\right)^{1/3}\,,
\end{equation}
where $M(r) = 4\pi\int^r_0{\rm{d}}~r'r'^2\delta\rho_{\rm m}(r')$ is the
mass enclosed within a radius $r$. This Vainshtein radius is an important quantity that provides the limit on the distance from the center of overdensity of $M(r)$ below which the fifth force is suppressed efficiently; the so called the Vainshtein screening mechanism \citep{Vainshtein:1972sx}.
 
In other words, if we consider the ratio of fifth force, $F_{\rm 5th}$ to Newtonian forces, $F_{\rm N}$:
\begin{equation}
\frac{F_{\rm 5th}}{F_{\rm N}} = \frac{\varphi_{,r}}{\Phi_{{\rm N},r}} \to 0\,,  \qquad {\rm as} \qquad \frac{r}{r_V} \to 0\,, 
\end{equation} 
the fifth force is suppressed near massive objects where $r << r_{V}$ and $\varphi_{,r}\to 0$, allowing the model to recover GR in high density regions.
 
The nature of Vainshtein mechanism makes difficult to
test such models in small scales in the non-linear regime. 
But constraints from the Solar System set an upper value of $[r_c H_{0}]^{-1} < 0.1$ \citep{Koyama2016}. On the
other hand, observations of luminous red galaxies from the BOSS-DR12 sample constrains the parameter to $[r_c H_0]^{-1} < 0.97$ \citep{Barreira:2016ovx}.

\begin{table*}
\centering
\begin{tabular}{c|c|c}
\hline
parameter & definition & value \\
\hline
$\Omega_m$  & present matter density & $0.281$ \\
$\Omega_{\Lambda}$ & $1-\Omega_m$ & $0.719$ \\
$h$ & $H_0/(100$~km~s$^{-1}$Mpc$^{-1})$ & $0.697$ \\
$n_s$ & primordial power spectral index & $0.971$ \\
$\sigma_{8}$ & r.m.s. linear density fluctuation & $0.820$ \\
\hline
$|f_{R0}|$ & Hu \& Sawicki $f(R)$ parameter &  $10^{-6}$ (F6), $10^{-5}$ (F5), $10^{-4}$ (F4) \\
$H_0r_c$ & nDGP parameter & $5$ (N5), $1$ (N1) \\
\hline
$L_{\rm box}$ & simulation box size & 1024~$h^{-1}$Mpc\\
$N_{\rm p}$ & simulation particle number & $1024^3$\\
$m_{\rm p}$ & simulation particle mass & $7.78\times 10^{10}h^{-1}M_{\odot}$\\
$N_{\rm dc}$ & domain grid cell number & $1024^3$\\
\hline
\end{tabular}
\caption{The cosmological parameters and specifications of the N-body simulations.}
\label{table:simulations_parameter}
\end{table*}

\section{$N-$body simulations}
\label{sec:N-body}

This Section describes the $N-$body simulations employed for this work. Here, we 
use six different $N-$body simulations corresponding to different gravity models, including the the standard $\Lambda$CDM one, based on GR. In this paper, we consider 
two different sets of MG models (as we explained above): the chameleon $f(R)$ gravity and the normal branch of the DGP braneworld model. 

All the simulations were generated using the adaptive-mesh-refinement {\sc amr} code {\sc ecosmog} \citep{Li2012,Li2013} based on the WMAP9 background cosmology \citep{WMAP9}:  $\Omega_{m0} = 0.281, h = 0.697$, and $n_s = 0.971 $. The simulation are in boxes of side length of 1024 $h^{-1}$Mpc, with $1024^3$ dark matter particles, and the respective particle mass of $m_p = 7.8 \times 10^{10}~h^{-1}M_\odot$, with a power spectrum normalization of $\sigma_{8} = 0.82$. Initial conditions 
for all the simulations were generated using the Zel'dovich approximation with the publicly available {\sc Mpgrafic} code \citep{Prunet2008} at $z_{\rm ini}=49$. We use the outputs at $z=0$.

The $f(R)$ and nDGP model parameters were chosen such that they deviate from $\Lambda$CDM only at later times. At $z_{\rm ini}$ MG effects can be neglected, and thus all simulations were run with the same initial condition. 
The maximal force resolution in all the simulations from the adaptive-mesh-refinement (AMR) technique is of $0.015~h^{-1}$ Mpc. This allow us to resolve over-dense regions where the screening effect is strong. 

For more details about the simulations, we refer the reader to \citet{Li2012} and \citet{Bose:2016wms}
for the $f(R)$ case, and \citet{Li2013} and \citet{Barreira:2015xvp} for the nDGP case.
The cosmological and technical parameters are given in Table~\ref{table:simulations_parameter}. At $z=0$, each simulation has five realizations which were run by slightly different initial conditions in the random phases of the density field. Except for F4 model where there are only 2 realizations available. We use these realizations to understand the statistical uncertainties in our results. 

Halos and subhalos where identified with the phase–space temporal 
halo finder \rockstar\footnote{https://bitbucket.org/gfcstanford/rockstar}  \citep{Behroozi2013}. Here, we use the halo mass definition of $M_{200c} \equiv \frac{4\pi}{3}200 \rho_{c}R^3_{200c}$ which corresponds to halos enclosing 200 times the critical density of the Universe, $\rho_{c}$ within the radius $R_{200c}$.
In all the six halo catalogs, we provide a limit in the number of particles contained in the halos, i.e  at-least the halo must contain 50 or more particles. This leads to the resolution limit in the maximum circular velocity of dark matter halos, $V_{\rm max}\sim 257$ km s$^{-1}$.

\subsection{Dark matter halo demographics}
\label{halo-demographics}


\begin{figure*}
     \centering
     \begin{tabular}{cc}
       \includegraphics[width=0.9\textwidth]{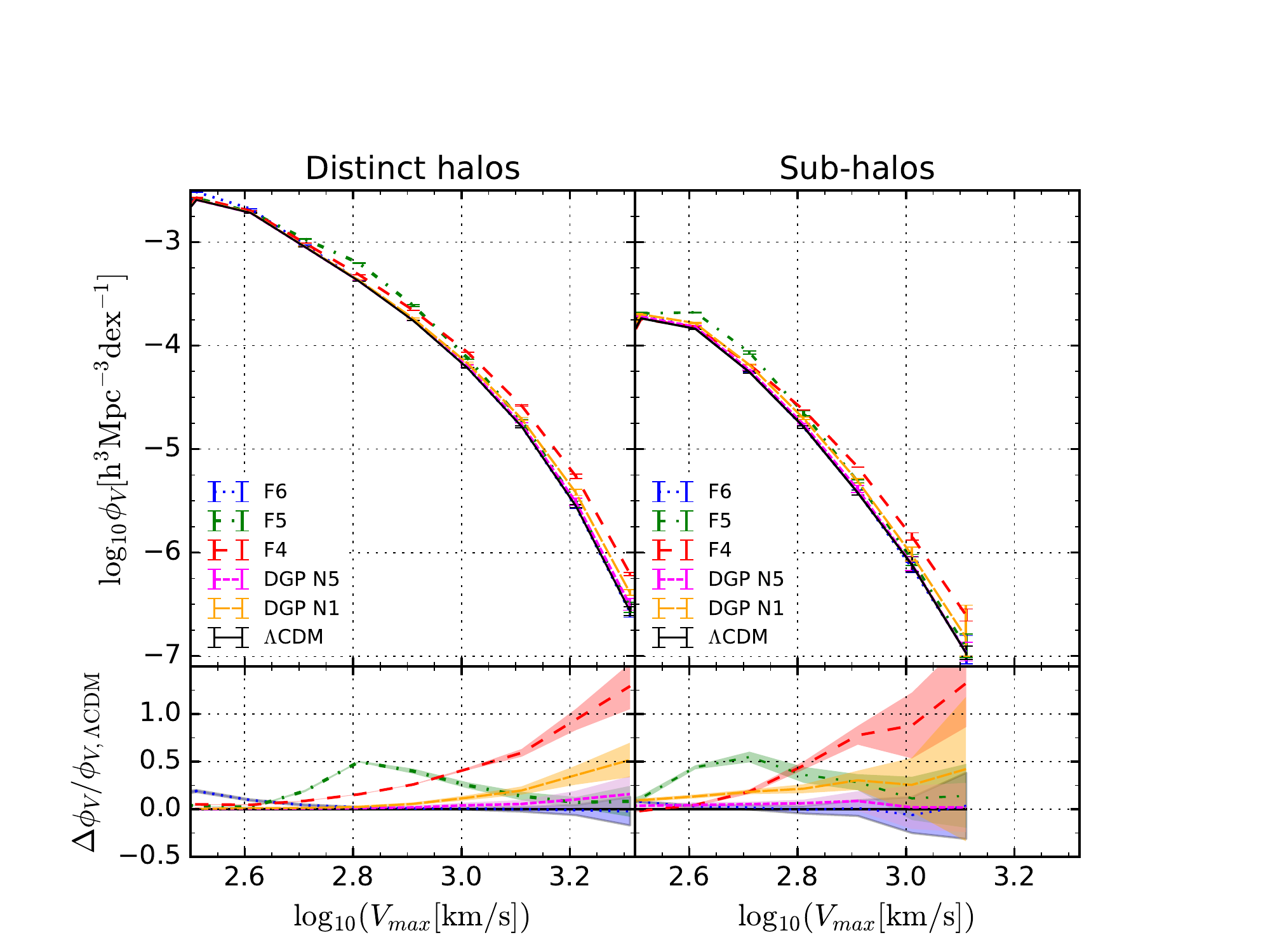}  
     \end{tabular}

\caption{Differential $V_{max}$ function of the distinct and sub halos for the MG models considered here along with the $\Lambda$CDM model. The black solid, blue dotted, green dot-dashed, red dashed, magenta medium-dashed, and orange long-dashed lines, represent $\Lambda$CDM(GR), F6, F5, F4, N5 and N1, respectively. This convention will be followed in the rest of the plots. The relative differences w.r.t the $\Lambda$CDM are shown in the lower panel, where the shaded regions show the error propagated from the one $\sigma$ standard deviations of all the realizations available. On comparing this figure with the corresponding halo mass functions and mass--concentration relations showed in Fig. \ref{fig:HMF}, a similar behavior is observed.
}  
\label{fig:Vmax}
\end{figure*} 

Figure \ref{fig:Vmax} shows the differential maximum circular velocity functions, $\phi_V( V_{\rm max})$, 
for distinct dark matter halos (left panel) and subhalos (right panel) for all the gravity models at 
redshift $z=0$. Here, subhalos are those halos, whose radius is inside of a larger halo; distinct halos can not be subhalos by definition. The black solid, blue dotted, green dash-dotted, red dashed, magenta medium dashed, and orange long dashed lines represent GR, F6, F5, F4, N5 and N1 models respectively. We will use this convention for all the figures representing these models throughout the paper.  The lower panel shows the relative difference w. r. t. the $\Lambda$CDM (GR) simulation, and the corresponding shaded areas represent the error propagated from the $1\sigma$ standard deviation measurement of all the realizations available as described above.
The differential maximum circular velocity function, $ \phi_V( V_{\rm max})$, shown in both panels are the mean of all the realizations for each gravity model analyzed in this paper.

We begin by describing the differences with the distinct halo velocity functions. 
As expected from theoretical arguments, see Section \ref{sec:Models_mod_grav}, we observe that 
in general, the smaller the value of the $f_{R0}$ parameter, the closer $\phi_V( V_{\rm max})$ to the standard $\Lambda$CDM model.

Based on the above, it is reasonable that the F4 model presents a significant difference from the standard $\Lambda$CDM model. This is also true when analyzing the nDGP models, especially at the high velocity-end both for distinct halos and sub-halos. 

Figure~\ref{fig:Vmax} has various point that
are worth to highlight. We focus first on $f(R)$ gravity models and 
divide our discussion into three different velocity ranges: 1) $ V_{\rm max}<500$ km s$^{-1}$; 2) 
$500$ km s$^{-1}$ $\leq V_{\rm max}<1000$ km s$^{-1}$; and 3) $1000$ km s$^{-1}$ $\geq V_{\rm max}$. The reason for the above three velocity ranges is that the behavior of the velocity functions for the $f(R)$ gravity models vary with the present day value of $f_{R0}$
as can be seen in the bottom panel of Figure~\ref{fig:Vmax}.
This panel shows that F6 has a maximum excess of abundance of halos of $\sim25\%$ w.r.t. the $\Lambda$CDM for halos with $ V_{\rm max}<500$ km s$^{-1}$ whereas both F5 and F4 remain roughly similar to the $\Lambda$CDM model. 
The recent analysis of high-resolution $N-$body simulations in \cite{Shi:2015aya} 
shows that the halo mass function of F6 has an excess $\sim20\%$ w.r.t. the $\Lambda$CDM for halos below $M_{200}\sim 10^{13} h^{-1}$ M$_{\odot}$, which is equivalent to our velocity range and thus consistent with our results. 
The authors interpreted the above as the result of the chameleon screen mechanism begins less efficient in those dark matter halos, see also Figure 8 in \cite{Falck:2015rsa}. 
Additionally, the authors showed that the halo profiles for those halos are significantly steeper in their inner regions than their GR counterparts. In consequence, their halo concentrations are enhanced and therefore their $V_{\rm max}$. This further 
explains why we observe the above difference between F6 and the $\Lambda$CDM model at these velocities.  

For halos with $500$ km s$^{-1}$ $\leq V_{\rm max}<1000$ km s$^{-1}$, 
we find an abrupt deviation for F5 w.r.t. $\Lambda$CDM of 
$50\%$. The difference peaks at $V_{\rm max} \sim 630$ km s$^{-1}$ but it decreases for both
low and high velocities. In the case of F4 there is a clear systematic 
increase in the abundance of halos, reaching an excess of $\sim50\%$ at  $V_{\rm max} \sim 1000$ km s$^{-1}$ while F6 remains very similar to the $\Lambda$CDM model. The above differences observed in F5 are also a direct consequence of the chameleon screen mechanism. As shown in \cite{Li_Efstathiou_2012}, a feature of the chameleon screen mechanism is to produce an excess of dark matter halos within some mass range. This leads to a compensation effect by decreasing the number of low-mass halos\footnote{We do not observe this decrease due to resolution limits in the suite of simulations we are using.} due to the hierarchical mass assembling process of dark matter halos. On the other hand, as these authors noted, in high-mass halos the chameleon screen mechanism is very efficient in addition that the fifth force is suppressed due to high mass halos living in dense environments; thus, the abundance of high mass dark matter halos approaches to the $\Lambda$CDM model. Indeed, we observe that this is the case for the F5 model. In Appendix \ref{app:HMF_cvir} we present the corresponding halo mass functions in the left panel of Figure~\ref{fig:HMF}. Here the excess in the F5 model is evident and consistent with the discussion from \cite{Li_Efstathiou_2012}. Additionally, we also observe that halo concentrations in the F5 model are significantly enhanced within the mass range of halo excess (right panel of Fig. \ref{fig:HMF}), consistent when extrapolating the result from \citet{Shi:2015aya}. 

In general, the mass range at which the chameleon screen mechanisms produce an excess 
of dark matter halos depends on the present day value of the scalaron 
field $f_{R0}$, see Figure 9 from \cite{Li_Efstathiou_2012}. 
The higher the value in $f_{R0}$, the larger the halo masses where the excess of dark matter halos happens and the larger this excess. 
Finally, the increase of dark matter halos for F4 at $V_{\rm max}\ga 1000$ km s$^{-1}$ is just a consequence of the above. We thus conclude that the trends we observe over the full velocity range is result of the chameleon screen mechanism and its effects depending on the present day value of the scalaron field $f_{R0}$. Similar results have been noted in previous works using $N-$body simulations \cite[see e.g.,][]{Li2012,Li2013,Hernandez-Aguayo:2018yrp}.

As for the nDGP models, we observe marginal deviations from the $\Lambda$CDM.
These deviations are in the expected direction as we observe that N1 has a excess of halos at $V_{\rm max} \ga 1000$ km s$^{-1}$ but N5 remains very similar to the $\Lambda$CDM model. Finally, in the case of the subhalos, we observe a similar behavior to distinct halos but with larger uncertainties, which makes difficult to conclude over the significance of the result. Nontheless, similar results have been reported in \citet{Shi:2015aya} for the F6 model.

The above differences show the effect of the different screening mechanisms over all ranges of $V_{\rm max}$. Due to the extreme nature of gravity in F4 and N1 models, the corresponding screening mechanism (chameleon or Vainshtein) is less efficient for massive halos, i.e. halos with larger $V_{\rm max}\gtrsim 1000$ km s$^{-1}$, leading to an increase in the number density of halos in these models. For F5, the inefficiency of the chameleon screening mechanism appears in halos with intermediate values of $V_{\rm max}\sim600$ km s$^{-1}$, as discussed above. On the other hand, the difference is small for F6 and N5 models w.r.t. $\Lambda$CDM.

\section{The Galaxy-Halo connection in Modified Gravity simulations}

\label{sec:galaxy-halo_connection}

To assign galaxies to the dark matter halos we use the (sub)halo abundance matching (SHAM). The match is between the SDSS $r-$band luminosity function and the $V_{\rm max}$ halo function, and then we obtain the the $M_r-V_{\rm max}$ relation. 
In this Section, we show that the $M_r-V_{\rm max}$ relation, for both central and satellite galaxies, varies among the models with differences below $\sim1\%$, which corresponds to differences of $\sim 0.2$ magnitudes at a fixed $V_{\rm max}$. We also show that the different models predict similar galaxy clustering, which are in good agreement with observations. 

\subsection{Subhalo Abundance Matching: SHAM}

Subhalo Abundance Matching, SHAM, is a simple rule that connects dark matter halo
properties to galaxy properties under the assumption that there is a 
one-to-one monotonic relation between galaxies and dark matter halo properties. In
addition, SHAM assumes that centrals and satellite galaxies have identical relationships. In this work we decide to use the maximum circular velocity of dark matter halos $V_{\rm max}$
as our halo property and $r-$band magnitudes for galaxy properties. 

The standard assumptions in SHAM have been criticized in previous studies 
\citep[see e.g.,][]{Rodriguez2012,Yang2012,Rodriguez2013}. 
These studies have shown that assuming similar relations for centrals and satellites
could lead to potential inconsistencies, particularly in reproducing galaxy clustering
and the observed counts from conditional stellar mass functions \citep[for a more
detail discussion see][]{Rodriguez2012}. Following \cite{Rodriguez2012}, we 
choose to separately derive relations for centrals and satellites as we will explain in more detail below. 

As mentioned above, in order to perform SHAM we consider to use the maximum circular 
velocity of halos, $V_{\rm max}$, 
as the main halo property that correlates with galaxy luminosities. Indeed, previous
studies \citep[e.g.,][]{Conroy+2006,Reddick+2012,Campbell+2017} have shown that when using 
$V_{\rm max}$ for dark matter halos, the galaxy clustering is in much better agreement
with observations. In particular, the above works have shown that using halo
$V_{\rm max}$ at $z\sim 0$ for distinct halos and the highest  $V_{\rm max}$ 
reached along the main progenitor branch of the halo’s merger tree, typically referred 
as $V_{\rm peak}$, for subhalos, give  much better results. This is because the properties 
of a halo’s central region, where 
central galaxies reside, are better described by $V_{\rm max}$ than halo mass. Unfortunately,
values for $V_{\rm peak}$ are not available in our set of simulations and
contrary to other works, here we use $V_{\rm max}$ for  both distinct and
subhalos. Note however, that \cite{Rodriguez2012} showed that using separate relationships
for central and satellites, the spatial galaxy clustering is well recovered even when employing only current ($z\sim0$) halo 
properties. 
Using a similar approach as in that paper, we will show that our 
mock galaxy catalogs are able to recover the observed spatial galaxy clustering. Finally, we mention that SHAM has been 
applied for $f(R)$ models 
in previous works \citep[see e.g.,][]{He:2016uxb}.

Next, we discuss the observational inputs that we use for the SHAM, that is, the SDSS $r-$band luminosity functions of central and satellite galaxies. 

\subsubsection{$r-$band Luminosity Functions}

Here, we use the SDSS DR7 $r-$band luminosity function measured in \cite{Dragomir2018}
for all galaxies. 
The authors used the New York Value Added Galaxy Catalog (NYU-VAGC; \citealp{Blanton2005b}) 
based on the SDSS DR7 which comprises a catalogue of $6\sim10^{5}$ spectroscopic galaxies 
over a solid angle of 7748 deg$^2$ in the redshift range $0.01<z<0.2$. The authors
K+E-corrected $r-$band absolute magnitudes at $z=0$. In order to derive the luminosity
function for centrals and satellites we used the results from \citet{Yang2009}. Using 
the NYU-VAGC based on the SDSS DR4, \citet{Yang2009} derive the 
$r-$band luminosity functions for centrals and satellites separately. Below,
we describe our best fit model to the fraction of satellite galaxies as a function
of $r-$band luminosity.

\begin{figure}
\begin{tabular}{cc}
\includegraphics[width=\columnwidth]{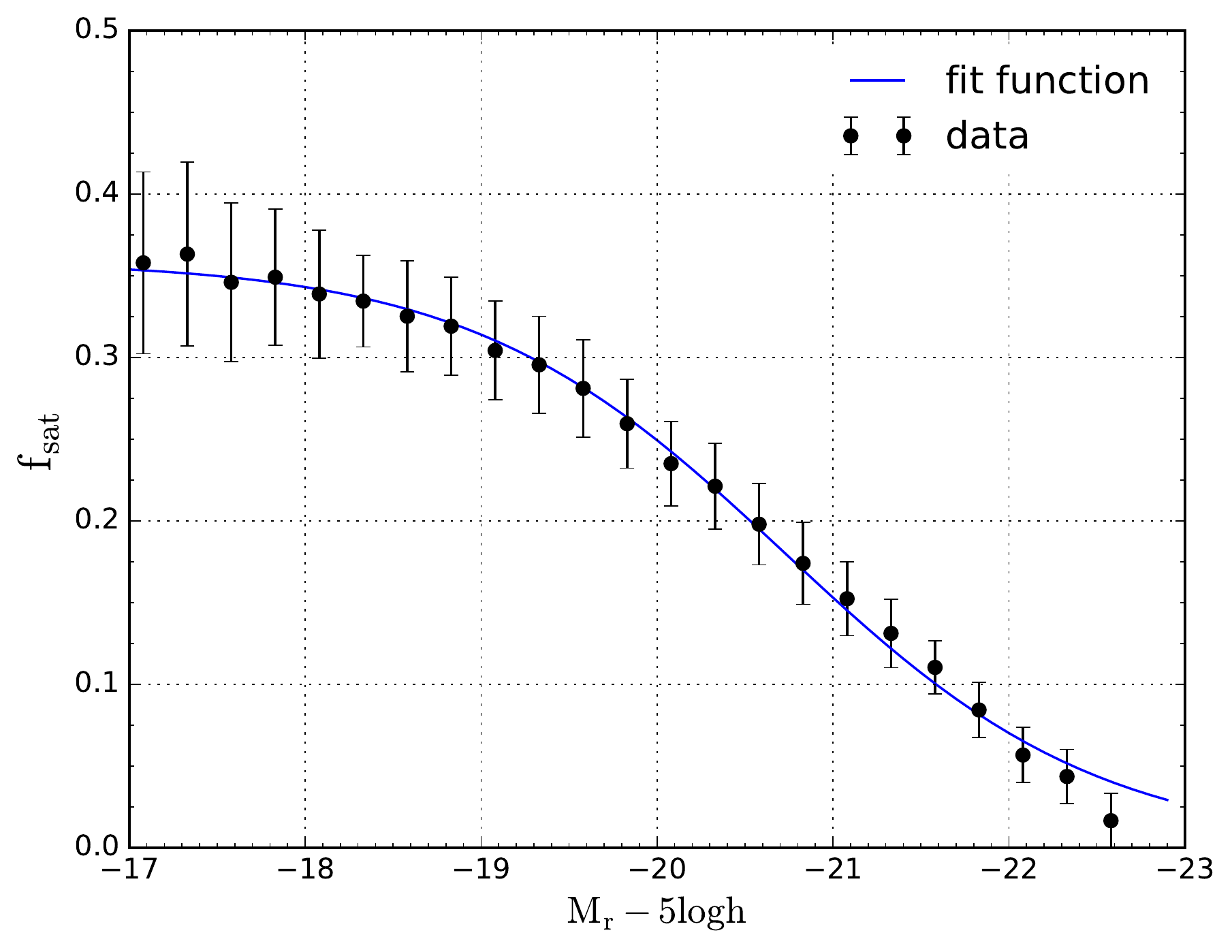}  
\end{tabular}
\caption{Satellite fraction as a function of $M_{r}-5\log h$. The black dot with error bars correspond to the fraction of satellites inferred by \citet{Yang2009} from the SDSS data, and the blue line is the best fit using the function Eq. (\eqref{eq:fraction_sat})}. 
\label{fig:sat_fraction}
\end{figure}

The fraction of satellite galaxies is defined as the ratio of the satellite luminosity function to the
total, $f_{\rm sat} = \phi_{\rm sat} / \phi_{\rm tot}$; the fraction of central
galaxies is simply $f_{\rm cen} = 1 - f_{\rm sat}$. After some experimentation with different functional forms, we find that the following function reproduces accurately the observations:
\begin{equation}
	f_{\rm sat}(M_{\rm r}) = \frac{A}{(1+10^{\alpha( M^{\ast}- M_{\rm r})})},
\label{eq:fraction_sat}
\end{equation}
where $A$ is an amplitude, $M^{\ast}$ is the characteristic magnitude 
at which the fraction $f_{\rm sat} = A/2$ and 
$\alpha$ controls the slope of the fraction at the massive-end. Note that in Eq.~\eqref{eq:fraction_sat} we use $r-$band absolute magnitudes $M_{r}$ while \citet{Yang2009}
reported luminosities. We use $\log L = - 0.4 \times( M_{\rm r} - M_{{\rm r}, \odot})$  with
$ M_{{\rm r}, \odot} = 4.67$. Figure \ref{fig:sat_fraction} shows the observed fraction 
of satellite galaxies as a function of $M_r$ magnitude, filled circles with error bars. 
We find that the best fitting values are $(A, M^{\ast},\alpha) = ( 0.357, -20.741, 0.504)$,
the best fit model is shown the blue solid line.

\begin{figure*}
     \centering
     \begin{tabular}{cc}
       \includegraphics[width=0.8\textwidth,angle=0]{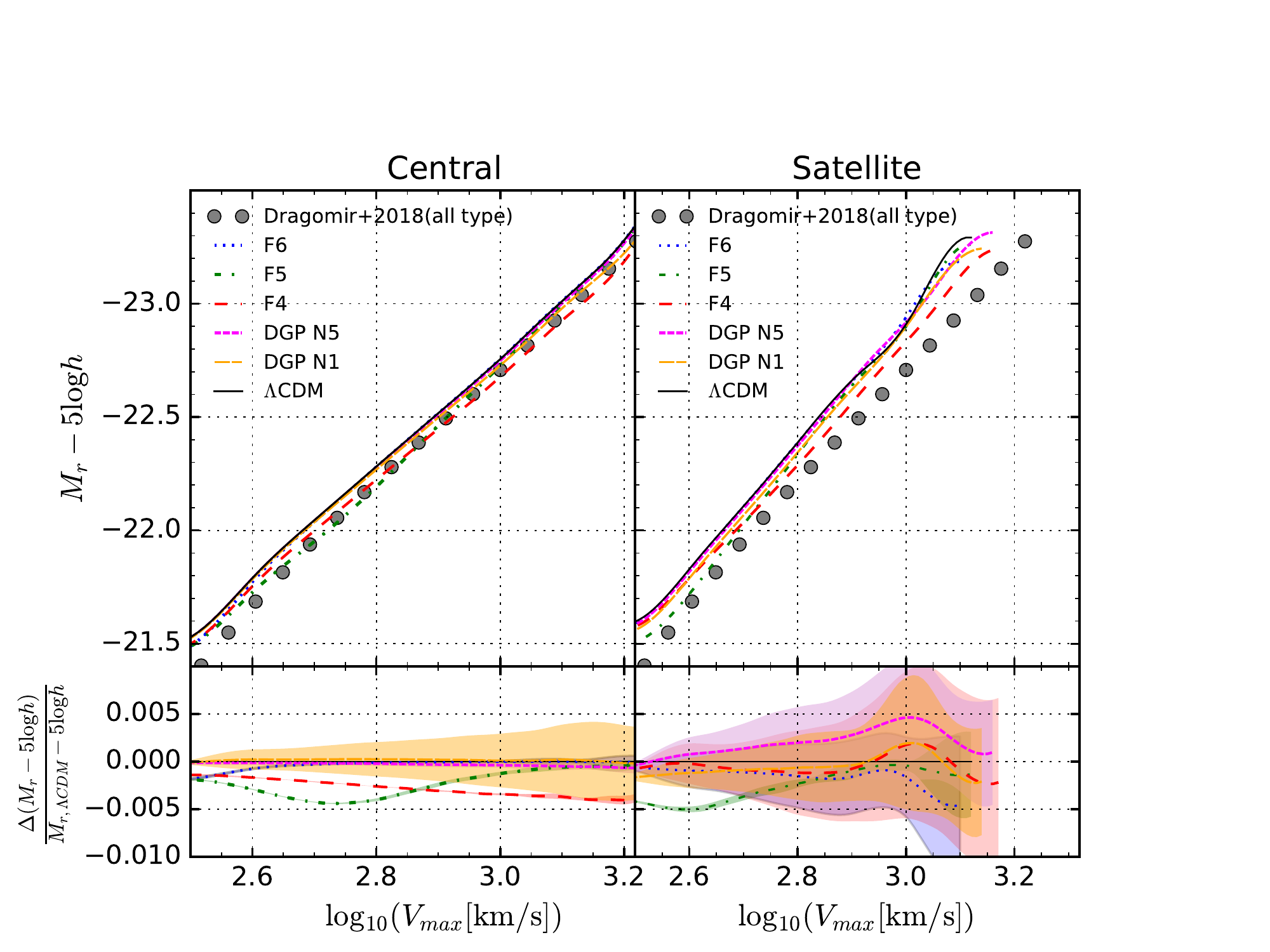}   & 
     \end{tabular}

\caption{Obtained central and satellite $M_{\rm r}- V_{max}$ relations by means of the SHAM method for our different gravity models. The relative differences w.r.t $\Lambda$CDM are shown in the lower panel. The result of \citep{Dragomir2018} for all type galaxies (see Appendix A of their paper) is reproduced in both panels with gray filled circles. Some differences in $M_{\rm r} - V_{max}$ relation are observed in all MG models with respect to $\Lambda$CDM but the variations are $<1\%$, mostly within the 0.5\%.} 
\label{fig:Magr}
\end{figure*} 

We use the above best fitting model to derive the $r-$band luminosity function
separately for centrals and satellites for SHAM. As mentioned above, we are considering the luminosity function from \citet{Dragomir2018}. Specifically, we use their best fitting model to 
a modified double Schechter function, see their equation (4), with their best fit parameters
reported in their Table 1.

\begin{figure*}
     \centering
     \begin{tabular}{cc}
       \includegraphics[width=0.8\textwidth,angle=0]{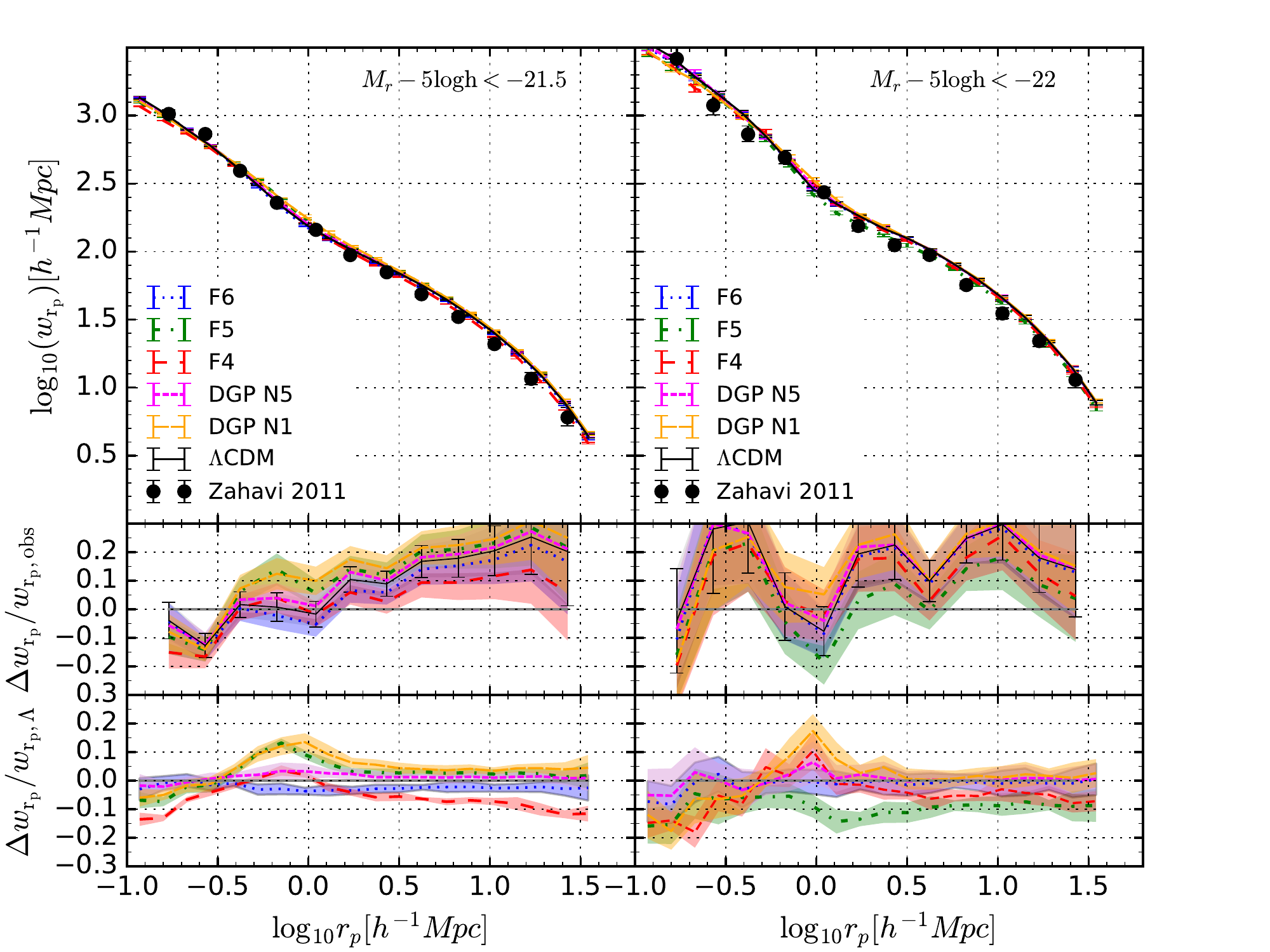}   & 
     \end{tabular}

\caption{{\it Upper Panels}: Projected two-point correlation function in the $r$-band calculated from the simulation-based mock catalogs of the different gravity models, using our SHAM results for the galaxy-halo connection. We split the catalogs into two sets of luminosity limits, $M_{r}-5\log h < -21.5$ and $M_{r}-5\log h < -22$ (left and right panels, respectively). The black circles with error bars show the results from SDSS DR7 \citep{Zehavi2011}. {\it Middle Panels}: Relative differences of the model predictions with respect to the observational data points. The black error bars show the 1-$\sigma$ error in the observational results propagated to the differences with respect to the $\Lambda$CDM model. The shaded color regions represent the variations estimated from the different realizations of each MG model. Note that though we observe differences up to $30\%$ in amplitude with respect to observations, the trends in the differences are similar for all the models, including the $\Lambda$CDM one. This means that the main differences with respect to observations are rather systematical for all the cases, and related mainly to the fact that we assumed no scatter in mocking the galaxies (see text). 
{\it Lower Panels}: Relative differences of the MG models with respect to the $\Lambda$CDM.  
Note that the main differences of MG models with respect to the $\Lambda$CDM one is mostly within $\sim 10\%$ in the left panel and 
$\sim 20\%$ in the right panel.}. 
\label{fig:Projected2points}
\end{figure*} 

\subsubsection{Results on the Galaxy-Halo Connection}

We apply SHAM separately for centrals and satellites to find the $M_{\rm r}-V_{\rm max}$ relation \citep{Rodriguez2012}. For centrals we use:  
\begin{equation}
\int_{M_{\rm r}}^{\infty} \phi_{\rm cen}(M_{\rm r}')d M_{\rm r}' = \int_{V_{\rm max}}^{\infty}\phi_{V,{\rm halos}}(V_{\rm max}') d\log V_{\rm max}',
\end{equation}
and similarly for satellites
\begin{equation}
\int_{M_{\rm r}}^{\infty} \phi_{\rm sat}(M_{\rm r}')d M_{\rm r}' = \int_{V_{\rm max}}^{\infty}\phi_{V,{\rm subhalos}}(V_{\rm max}') d\log V_{\rm max}'.
\end{equation}
Note that the above form of SHAM implies zero-scatter in the $M_{\rm r}-V_{\rm max}$ relations.

Figure \ref{fig:Magr} shows respectively the resulting $M_{\rm r}-V_{\rm max}$ of centrals and
satellites in the left and right upper panels for the models considered respectively. 
We also compared to the results from \citet{Dragomir2018}, who applied the same SHAM we use here. \citet{Dragomir2018} derived the above relation for all type galaxies; as discussed in \cite{Rodriguez2012} the relation for all galaxies is closer to the one of centrals than for satellites. Differences with \cite{Dragomir2018} is just mainly the result of different cosmological parameters: these authors used a cosmology based on the results from Planck 2016 \citep{Ade2015}, while here we use a cosmology that is closer to the results from the WMAP9 mission. \footnote{For Planck 2016, the total matter density at the present day is $\Omega_{m}\approx 0.308$ whereas for WMAP9, this density is lower, $\Omega_{m}\approx 0.28$.} One can see in figures 10 and 15 of \citet{Rodriguez+2016a} the ratio of the predicted distinct halo number densities between the Bolshoi-Planck and Bolshoi-WMAP7 ($\Omega_m = 0.27$, similar to WMAP9) simulations; similar differences are observed by us. Due to this difference, a slightly higher number density of dark matter halos at a fixed halo mass/$V_{\rm max}$ are found in the Bolshoi-Planck simulation than in our simulation. The SHAM translates this into a shift to higher values of $V_{\rm max}$ for a given luminosity. Note also that the SHAM in \citet{Dragomir2018} was applied to all halos while here we separate into distinct halos and subhalos. 

The lower panels in \ref{fig:Magr} present the differences in the $M_{r}$ magnitudes
at a fixed $V_{\rm max}$ for all the MG models with respect to the standard $\Lambda$CDM model. The shaded regions show the $1\sigma$ standard deviation calculated using 5 realizations 
except for F4, for which we consider only 2 realizations. Observe that these differences
are just a direct result of the differences between the velocity functions described in the previous subsection (see Fig. \ref{fig:Vmax}), and not to different cosmological parameters since all of our simulations use the same cosmological parameters.
Figure~\ref{fig:Magr} shows that the differences in $M_r$ for the MG models with respect to the 
$\Lambda$CDM (GR) one is not larger than $1\%$. This statement is valid for both central and satellite galaxies. 

Indeed, in both cases the differences in $M_r$ for all the models is within $0.5\%$. 
The above is especially true for F4 and F5 MG models. Note that N1 and N5 models are closer 
to the $\Lambda$CDM for the particular case of central galaxies. Note that these differences are not related to
uncertainties in the determination of the $M_r-V_{\rm max}$ relationship but true deviations due to the different gravity models. As noted above they resemble the differences in the velocity functions. 
In terms of magnitudes the differences are up to $\sim 0.2$ mags. We do not propagate 
uncertainties due to observations as it will be the same for all the models. Next, we discuss errors in the $M_r-V_{\rm max}$ relationship determinations. 

We measure uncertainties around the $M_r-V_{\rm max}$ relationships by using the various realizations from our suite of simulations, shown as shaded areas in the lower panels of Figure \ref{fig:Magr}.
This figure shows that for central galaxies N1 has the largest error when determining the
$M_r-V_{\rm max}$ relationship over all realizations. However, it is not larger than $\sim0.5\%$, see orange shaded area.
In the case of satellite galaxies, their number is much lower than centrals and thus more dominated by Poissonian error. While we expect larger uncertainties for satellites, we observe uncertainties not larger than $\sim1\%$ in most of the models. 
The fact that uncertainties, both for central and satellites, are smaller than $\sim1\%$ in the $M_r-V_{max}$ relationship guarantees accurate mock galaxy catalogs, and that cosmic variance is not an extra uncertainty in our determinations. 
In other words, {\it differences in our resulting predictions based on the mock catalogs will be the result of the differences among the different gravity models rather than in the technique itself or from cosmic variance.} We will come back to this in Section \ref{secc:discusion}. 

\subsection{Spatial Galaxy Clustering}

As a standard procedure, we show that our resulting $M_{\rm r}-V_{\rm max}$ relations for
centrals and satellites is consistent with the observed projected two-point correlation functions from the SDSS DR7 \citep{Zehavi2011}. Indeed, spatial galaxy clustering should be considered as a consistency test for the mock galaxy catalogs that we generated for all the MG models. 

\citet{Zehavi2011} derived the projected two-point correlation function in the $r-$band magnitude at $z=0.1$ for several magnitude thresholds. Recall that in this paper we are using
the results from \citet{Dragomir2018}, who used $r-$band magnitude at $z=0$. In order
to make a fair comparison between our predicted and the projected two-point correlation
from \citet{Zehavi2011}, we transform our $r-$band magnitudes to $z=0.1$ using the 
relationship reported in \citet{Dragomir2018}. 

Figure \ref{fig:Projected2points} shows the projected two-point correlation function
for two magnitude thresholds; $M_{r} -5 \log {\rm h} < -21.5$ and $M_{r} -5 \log {\rm h} < -22$ in left and right panels, respectively.
The middle and bottom panels show the relative differences of the different gravity models with respect to observations from the SDSS DR7, and of the MG models with respect to the $\Lambda$CDM (GR) model, respectively. The black error bars (showed only for the $\Lambda$CDM model) correspond to the uncertainties propagated from the observations, and the shaded areas show the $1\sigma$ error estimated using the realizations available for each gravity model.

There are various points to highlight from this figure. First, note that 
SHAM reproduces the projected two-point correlation functions for $M_{r} -5 \log {\rm h} < -21.5$ ($M_{r} -5 \log {\rm h} < -22$) within the $\sim20\%$ ($\sim30\%$) for all gravity models, see the middle panels of Figure \ref{fig:Projected2points}. 
As mentioned above the shaded areas show the dispersion from all the realizations of the simulations employed in this paper. Within the uncertainties, all the simulations are recovering acceptable correlation functions. Note, however, that 
there are systematic trends as a function of the projected radius that are independent of the gravity models employed in this paper. 
This implies that the differences from observations as a function of radius are not particularly intrinsic to some type of gravity model but a systematic arises when assigning galaxies to halos. Is this a sign that SHAM fails in reproducing
galaxy clustering even in the case of the standard $\Lambda$CDM? There are 
several understandable reasons why our models do not reproduce to a much higher accuracy galaxy clustering. 
The most obvious one is that we have assumed no scatter in the $M_r-V_{\rm max}$ relation both for centrals and satellites. The impact of including scatter reduces galaxy clustering at large projected distances as shown in \citet{Reddick+2012}. 
Even a moderate value of scatter, $\sim0.2$ dex, reduces galaxy clustering for large
distances but does not affect significantly the clustering at small distances, see
figure 5 from \citet{Reddick+2012}. This will simply explain why our galaxy mock models 
tend to overestimate more the two halo-term (larger distances) than the one
halo term in both panels. Since all the gravity models follow the same 
trend in the difference with the observed galaxy clustering, we do not consider this difference as an extra source of uncertainty for our further comparative study among the different gravity models. 

In more detail, we observe that the degree of agreement with respect to the observation as a function of the scale may be slightly different for each gravity model. For the magnitude limit of $M_{r} -5 \log h < -21.5$, the $\Lambda$CDM model, on average, remains within $\sim10 \%$ below $r_p \sim 2$ Mpc$/h$ and, on average, it deviates around $\sim 20\%$ above $r_p \sim 10$ Mpc$/h$. Surprisingly, F4 is on average closer to the observations, within $\sim10\%$ at all scales; recall that for this model, its  velocity function significantly deviates from the $\Lambda$CDM one. For the magnitude limit of $M_{r} -5 \log h  < -22$, note that uncertainties become larger both in observations as well as in the theoretical prediction of projected correlation functions. The reason behind it is quite simple as the number of brighter galaxies decreases both in the SDSS DR7 survey and in our mock galaxy catalogs. 
We expect that with on-going and future surveys, as well as with larger N-body simulations, the measurements of the galaxy clustering will improve further and will be sensitive enough to discriminate among different MG models. Here, despite of the differences described above it becomes difficult to conclude which MG model describes better the current observed galaxy clustering. Additionally, recall that our SHAM does not include scatter around the $M_r-V_{\rm max}$ relation. Even more, this scatter could be different for the different gravity models. In any case, the aim of this work is not to use galaxy clustering as a discriminant of gravity models. We just wanted to show that the SHAM applied to each one of the gravity models predicts spatial galaxy clustering in rough agreement with observations, and that differences in the clustering of the MG models with respect to the $\Lambda$CDM are mostly within the 10\%.

\begin{figure}
\begin{tabular}{cc}
\includegraphics[width=\columnwidth]{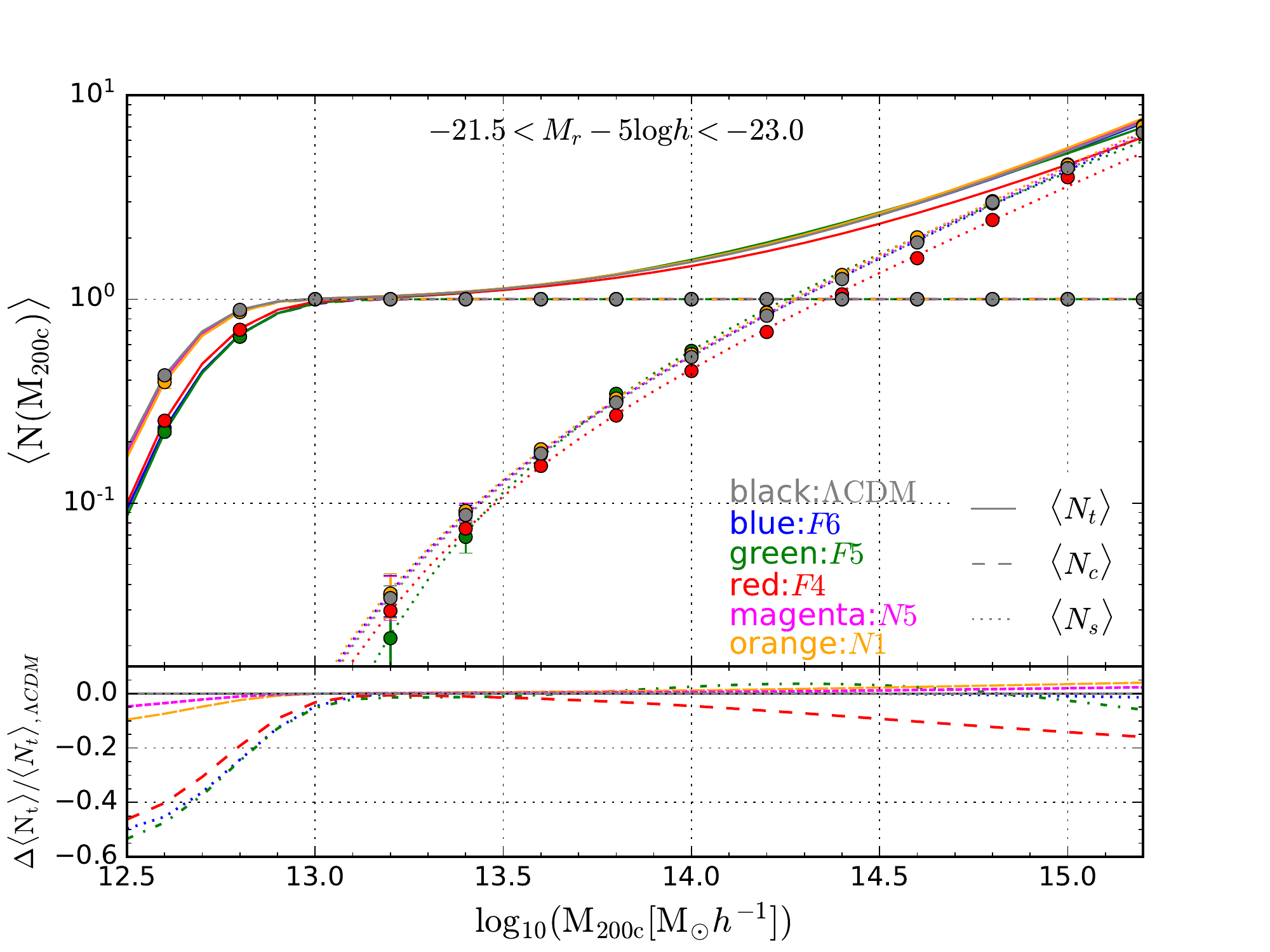}  
\end{tabular}
\caption{{\it Upper panel}: Mean galaxy occupation numbers as a function of host halo mass for central and  satellite galaxies (filled circles) as estimated from our mock galaxy catalogs for the $\Lambda$CDM and MG models; the color coding is shown inside the panel and it is the same as in Fig. \ref{fig:Vmax}. The dashed and dotted lines are best fits using the functional forms of mean occupation numbers for centrals (Eq.~\ref{eq:HOD_cen}) and satellites (Eq.~(\ref{eq:sat_hod}), respectively. The solid lines correspond to the total mean occupation numbers from the fit (Eq.\ref{eq:hod_t}). 
{\it Lower panel}: Relative differences of the fitted total mean occupation numbers between each MG model w. r. t. the $\Lambda$CDM. A sharp difference of $10-50\%$ is observed for the $f(R)$ models in the HOD corresponding to central galaxies in the mass range $10^{12.5} M_{\odot}h^{-1} < M_{200c}< 10^{13} M_{\odot}h^{-1}$. For the nDGP models, the differences remain within $10\%$ for central occupation numbers and within $\sim$5\% for satellite numbers. Similar differences in the satellite occupation numbers are seen for the $f(R)$ models, except for F4.}
\label{fig:HOD}
\end{figure}

\begin{table*}
\centering
\begin{tabular}{c c c c c c}
\hline
\hline
Models & ${\rm log} M_{1}$ & ${\rm log}M_{cut}$ & $ \alpha$ & $ {\rm log}M_{min}$& $ \sigma_{\rm log M}$ \\
\hline
$\Lambda$CDM    &$14.204\pm 0.012$ &$13.403 \pm 0.02 $  &$0.797\pm 0.021$ &$12.627\pm 0.0002  $ & $ 0.201\pm 0.001 $  \\
 F6            &$14.184\pm 0.014$ &$13.459 \pm 0.017$ &$0.739 \pm 0.027$ &$12.725\pm0.003    $ & $ 0.24\pm 0.006 $ \\
F5             &$14.156\pm 0.017$ &$13.508 \pm 0.018$ &$0.762 \pm 0.023$ &$12.727\pm0.003      $ & $ 0.236 \pm0.006  $  \\
F4             &$14.300\pm 0.011$ &$13.381 \pm 0.013$ &$0.79 \pm 0.020$ & $12.707\pm 0.002 $ & $ 0.228\pm 0.004 $ \\
N1             &$14.195\pm 0.01$ &$13.379 \pm 0.023$ & $0.80\pm 0.023$ & $12.64\pm 0.0003 $ & $ 0.205\pm 0.001$  \\
N5             &$14.198\pm 0.01 $ &$13.402 \pm 0.021$  & $0.80\pm0.021 $ & $12.633\pm0.0002 $ & $ 0.203\pm0.001   $ \\
\hline
\end{tabular}
\caption{Best fit values of the HOD parameters for satellite and central galaxies fitted to functions Eqs.~\eqref{eq:sat_hod} and \eqref{eq:HOD_cen}, respectively, based on the mock catalogs generated for all the models. Mass parameters are in units of $M_{\odot}h^{-1}$. Error bars on the HOD parameters correspond to $1\sigma$, derived from the marginalized distributions. The HOD parameters vary from model to model; the largest differences among the constrained parameters are:  $\sim 40\%$ between F4 and F5 for $M_{1}$; $\sim 35\%$ between F5 and N1 for $M_{\rm cut}$;  $\sim 26\%$  between $\Lambda$CDM and F5 for $M_{\rm min}$; $10\%$ between F6 and N1 for $\alpha$; and $20\%$ between $\Lambda$CDM and F5 for $\sigma_{\log M}$.
}
\label{table:bestfits}
\end{table*}

\subsection{Halo Occupation Distribution Model: HOD}
\label{sec:HOD_parameters}

In the context of SHAM, the halo occupation distribution of galaxies in dark matter halos
arises naturally. In this section, we explore and quantify the resulting 
halo occupation distribution (HOD) based on SHAM for our set of different gravity models. 
So far, HOD remains as one of the most powerful tools to connect the galaxies to the cold dark matter halos, 
with the aim to constraint various physical processes that govern the galaxies formation and 
its evolution \citep[for a review see,][]{Wechsler2018}. 
Nonetheless, HOD is not only a powerful tool for constraining galaxy formation but it is also a useful tool to constrain cosmology \citep[see e.g.,][and references therein]{Yang+2003,vandenBosch2012,More2012xa, More2012}.
Thus, it is interesting to study the changes in HOD under different cosmological scenarios 
specially when applying SHAM results in different $M_{r}-V_{\rm max}$ relations. 

Briefly, the HOD quantifies the probability of finding $N$ galaxies above some magnitude threshold within a given halo with mass $M_{200c}$. 
Here, we calculate the mean occupation of central and satellite galaxies above the limit of $M_{r} -5 \log h  < -21.5$.
Fig.~\ref{fig:HOD} shows the resulting HOD separately for centrals and satellites. 
Next we describe the best fitting functions to our HOD models  
and provide separate fits to the satellite and central mean occupation functions for all the gravity models studied in this paper. 
For the mean occupation of satellites, we use the standard power-law with an exponential decreasing function that consists of three free parameters \citep{Kravtsov+2004}:
\begin{equation}
\langle N_{\rm sat}(M_{200c}) \rangle = \left(\frac{M_{200c}}{M_{1}}\right)^{\alpha} \exp\left(-\frac{M_{\rm cut}}{M_{200c}}\right),
\label{eq:sat_hod}
\end{equation}
where the parameter $M_{1}$ corresponds to the mass scale at which halos host at-least a satellite on average, $M_{cut}$, the mass scale below which the satellite mean occupation starts to decay exponentially and $\alpha$ provides the power law slope of the relation between halo mass $M_{200c}$ and $N_{sat}$.
On the other hand, for the mean occupation function of central galaxies, we consider the standard error function representation:
\begin{equation}
\langle N_{\rm cen}(M_{200c}) \rangle = \frac{1}{2}\left(1+{\rm erf}\left(\frac{{\rm log } M_{200c}-{\rm log} M_{\rm min}}{\sigma_{\rm logM}}\right)\right),
\label{eq:HOD_cen} 
\end{equation}
where erf is an error function ${\rm erf}(x) = \frac{2}{\sqrt{\pi}}\int_{0}^{x} \exp(-t^2) dt$. The 
$ M_{\rm min}$ is the minimum mass at which a halo hosts at least one central galaxy and $\sigma_{\rm logM}$ width of the transition between ${N_{\rm cen}=0}$ and ${N_{\rm cen}=1}$. Under the assumption that the HODs of central and satellite galaxies are independent, we can fit the total mean occupation with 5 free parameters as
\begin{equation}
\langle N_{\rm tot} \rangle =  \langle N_{\rm cen} \rangle + \langle N_{\rm sat} \rangle.
\label{eq:hod_t}
\end{equation}
The resulting best fitting parameters for all the gravity models are shown in Figure \ref{fig:HOD} with the solid lines for centrals and dotted
lines for satellites. The color code corresponds to the different MG models using the above equations.
We report the best fit values of HOD parameters in Table \ref{table:bestfits}.
The maximum variation between the best fit values of the HOD parameters among the models are around $\sim40\%$ for $M_{1}$;
$\sim 35\%$ for $M_{\rm cut}$; $\sim 26\%$ for $M_{\rm min}$; $10\%$ for $\alpha$; and $20\%$ for $\sigma_{\log M}$.

The bottom panel of Figure \ref{fig:HOD} shows the differences w.r.t. the $\Lambda$CDM standard model as a function of halo mass. Here we observe that in the case of central galaxies, there is a difference around $10-50\%$ between MG models and $\Lambda$CDM, specially for $f(R)$ models at lower halo mass limits, $10^{12.5} M_{\odot}h^{-1} < M_{200c}< 10^{13} M_{\odot}h^{-1} $. As for N5 and N1 models they are distinguishable from
the standard $\Lambda$CDM in less than $5\%$ to $10\%$ at mass $10^{12.5} M_{\odot}h^{-1} $ respectively. The above results are not surprising as similar difference were noted in $M_{r}-V_{\rm max}$ relation for central galaxies in
Fig.~\ref{fig:Magr}. There, we noted that larger differences with respect to $\Lambda$CDM were observed in $f(R)$ gravity models than in nDGP models. So, the results of HOD
we have obtained are consistent with what we expected. The HOD of satellite galaxies is also different and depends on the gravity model
that is used. Recall, that in our case we are using subhalos as tracers of satellite galaxies as we have constrained separately the galaxy-halo
connection for satellite galaxies. At high mass limits, where satellite mean occupation number is dominated,
we find that the total mean occupation numbers of galaxies of F4 model show a deficit of $10-20\%$ from the $\Lambda$CDM model at mass $10^{14.5}- 10^{15} M_{\odot}h^{-1} $while rest of the MG models remains within $5\%$ difference. 

The above results lead to a clear conclusion: {\it 
{\rm HOD} parameters depend on the gravity model employed}. Thus, the use of HOD parameters derived from the $\Lambda$CDM cosmology and employed in MG simulations, would lead to {\rm biased} conclusions from the analysis of the mean occupation of galaxies, as one can clearly see from Figure \ref{fig:HOD}. This will potentially reflects on the inferred galaxy clustering in MG models that
will be in tension with observations.
In other words, the differences in the HOD models described in the above figure should be interpreted as the result of the same degree of success in the galaxy clustering
discussed in Figure \ref{fig:Projected2points}. 

{\it Independent constraints on the HOD of centrals and satellites may be an interesting possibility to search for some signature of MG. } 

\subsection{Measuring the dependence of the luminosity function on environment}
\label{Method_GLFs_enviroment}

As discussed in this paper, it is natural to expect some deviations from different statistics between gravity models. The most obvious is the demographics of dark matter halos, namely, the velocity functions described in Section \ref{sec:N-body}.
Indeed, as shown in Fig.~\ref{fig:Vmax}, we observe differences of approximately $\sim 50\%$ between the velocity functions 
$\phi_V (V_{\rm max})$ of the different MG models analyzed in the paper.  Unfortunately, direct
measurements of the velocity function from observations is not yet possible, thus, in this paper we 
look for other observable quantities that are natural projections of the halo velocity functions. The luminosity
function is one of the most obvious observable distribution. 
While the observed total luminosity function is, by construction, the same in all our gravity models, so we expect that the observed differences in the halo velocity functions affect the dependence of the luminosity function on environment.

Thus, in this section we attempt to understand the effect 
of MG under different density environments; 
ranging from under-dense regions like voids to highly dense regions like galaxy groups and clusters.

In the past, studies have already shown that the SHAM technique recovers the correct dependence of environmental density under the assumption of a $\Lambda$CDM universe \citep[see e.g.,][and references therein]{Dragomir2018}. In particular, the recent work of \cite{Dragomir2018} showed
that SHAM reproduces the correct dependence of the $r$-band galaxies luminosity functions for centrals and satellites of the SDSS DR7 from the Bolshoi-Planck simulation 
\citep{Klypin+2016,Rodriguez+2016}. Here we extend the study by \cite{Dragomir2018} to other gravity models to understand whether the observed dependence of the
luminosity function with environment is a simply tool to constrain gravity models. 

In order to quantify environments that can be directly compared to observations, particularly to volume-limited samples, 
we define a {\it density-defining population} or DDP \citep{Croton+2005}. Due to resolution limitations, for this work we define
DDP galaxies within the absolute magnitude limit of $-21.5 > M_{r} - 5 \log h > -22.5$. Our definition for the
DDP population is a compromise between the lowest masses sample in our mock simulations and observations that can be
done with current {\it galaxy surveys}, such as SDSS.

Among various existing methods to define local density environments \cite[see e.g.,][for a review]{Muldrew+2012}, 
we adopt the aperture-based methods where one measures the over-densities by counting the number of DDP galaxy neighbors, $N_{n}$, around each galaxy in the sample. Here, we are using an aperture of spheres of radius, $R_8 =8 h^{-1}$ Mpc and define the local density as
\begin{equation}
\rho_{R} =\frac{N_n}{4/3\pi r_{R}^3}.
\end{equation}
In order to calculate the density contrast, we need to determine the mean number density of galaxies, ${\bar {\rho}}$. 
Based on the global $r$-band luminosity function we find that the number density at the range $-21.5 >  M_{r} - 5 \log h > -22.5$ is ${\bar {\rho}}  = 5.824 \times 10^{-4}h^{3}$ Mpc$^{-3}$. Then, for every galaxy in the mock we measure the density contrast 
within a sphere of radius $R$ as:
\begin{equation}
\delta_{R} = \frac{\rho_R-{\bar {\rho}}}{\bar {\rho}}.
\label{eq:delta_g}
\end{equation}

Following previous studies \citep{Croton+2005,McNaught-Roberts+2015,Dragomir2018}, we focus our analysis mainly on  spheres of radius $R= 8 h^{-1}$ Mpc. This radius is optimal for sampling both under-dense and over-dense regions. Nonetheless, we also use a large search radii, $R=10h^{-1}$ Mpc, to
amplify the signal in void regions. We expected that voids contain more information about different gravity models due to
the nature of the screening mechanisms implemented on such MG models. 

In Section \ref{sec:results_section} we will also discuss results based on $R= 10 h^{-1}$ Mpc. 

Next, we describe the procedure to determine the galaxy luminosity functions for different environments. Here we 
follow \citet{Dragomir2018} for determining the effective volume,$\mathcal{V}_{\rm eff}(\delta_{R})$ as 
the fraction of effective volume $f(\delta_{R})$ sampled by the different environments in the simulation.
In other words, the effective volume is given by $\mathcal{V}_{\rm eff}(\delta_{R}) = f(\delta_{R})\times V_{\rm sim}$ 
where $V_{\rm sim}$ is the volume of the simulation. Hence, the $r$-band GLFs within a magnitude range of $M_{r} \pm \Delta M_{r}/2$ and $\delta_{R}\pm \Delta \delta_{R}/2$ density binning is given by:
\begin{equation}
\phi_{r}(M_{r},\delta_{R}) = \sum_{i =1}^{N} \frac{w_{i}(M_{r}\pm \Delta M_{r}/2, \delta_{R}\pm \Delta \delta_{R}/2)}{f(\delta_{R})\times \Delta M_{r}\times {V}_{\rm sim}},
\label{eq:phi_fraction}
\end{equation}
where $w_{i}$: 
\begin{equation}
w_{i} = \Bigg\{ \begin{array}{cc}
       \displaystyle 1 &
       \mbox{if $\delta_{r,i} \in (\delta_{r}\pm \Delta\delta_{r}/2)$ and $M_{r} \in  (M_{r}\pm \Delta M_{r}/2$)}\\
       0 & \mbox{Otherwise}
       \end{array}.
\end{equation}

Following \citet{Dragomir2018}, the fraction of effective volume, $f(\delta_{R})$, as function of $\delta_R$ is measured 
by counting the number of DDP galaxy neighbors around a catalog of random points:
\begin{equation}
f(\delta_{R}) = \frac{1}{N_{r}}\sum^{N_{r}}_{i =1} \Theta(\delta_{R,i}),
\label{eq:f_delta}
\end{equation}
where the function $\Theta$ counts the number of random points within the overdensity bin $\delta_{R}\pm \Delta \delta_{R}/2$ :
\begin{equation}
\Theta(\delta_{R,i}) = \Bigg\{ \begin{array}{cc}
       \displaystyle 1 &
       \mbox{if $\delta_{r,i} \in (\delta_{r}\pm \Delta\delta_{r}/2)$ }\\
       0 & \mbox{Otherwise}
       \end{array}.
\label{eq:step}       
\end{equation}
Similar to eq.(\ref{eq:delta_g}), the local density contrast around each random point $\delta_{r}$ is calculated as
\begin{equation}
\delta_{r} = \frac{\rho_r-{\bar {\rho}}}{\bar {\rho}}.
\label{eq:delta_random}
\end{equation}
We use a random catalog that is $\sim4$ times larger than the mock galaxy catalogs we created for all the gravity models, 
$\sim 4 \times 10^6$. We found that the variation in the fraction of effective volume measurements between the models with respect to $\Lambda$CDM is approximately within
$\sim 2\%$.

In the following we present results in six density bins:$ -0.4 < \delta_{8} < 0.0$, $ 0.0 < \delta_{8} < 0.7$, $ 0.7 < \delta_{8} < 1.6$, $ 1.6 < \delta_{8} < 2.8$, $ 2.8 < \delta_{8} < 4.0$ and $4.0 <\delta_{8}$.

\begin{figure}
\begin{tabular}{cc}
\includegraphics[width=\columnwidth]{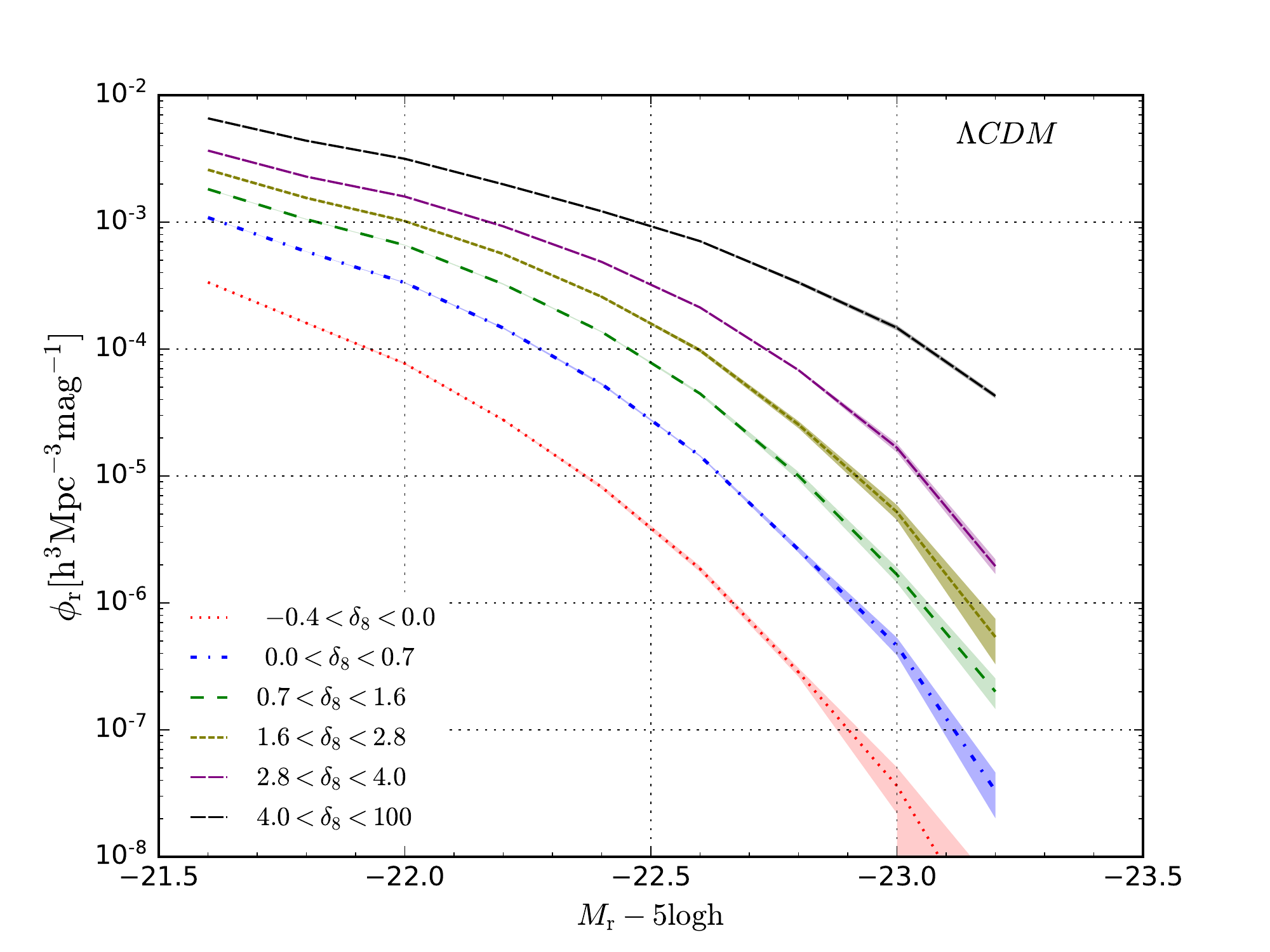}  
\end{tabular}
\caption{Galaxy luminosity functions in the $r$-band of all galaxies estimated from our mock galaxy catalogs for the $\Lambda$CDM model in six different density environments indicated inside the panel. Shaded regions represent 1 $\sigma$ standard  deviation measured from their 5 realizations available.}
\label{fig:lum_LCDM}
\end{figure}

Finally, Fig.~\ref{fig:lum_LCDM} shows the resulting dependence of the luminosity function on environment 
for the standard $\Lambda$CDM model. Note that as higher is the density environment bin the larger is
the number of galaxies per comoving volume. Additionally, the galaxy luminosity function
resembles a Schechter function, similarly to previous determinations for the $\Lambda$CDM model
\citep{Dragomir2018}.
While not shown here, we derive similar luminosity functions
for all the gravity models. In the next section we will show just the differences with respect 
to the $\Lambda$CDM model. 

\section{Galaxy distributions as a function of environment in the different gravity models}     
\label{sec:results_section}

\begin{figure*}
     \centering
     \begin{tabular}{cc}
        \includegraphics[width=0.5\textwidth,angle=0]{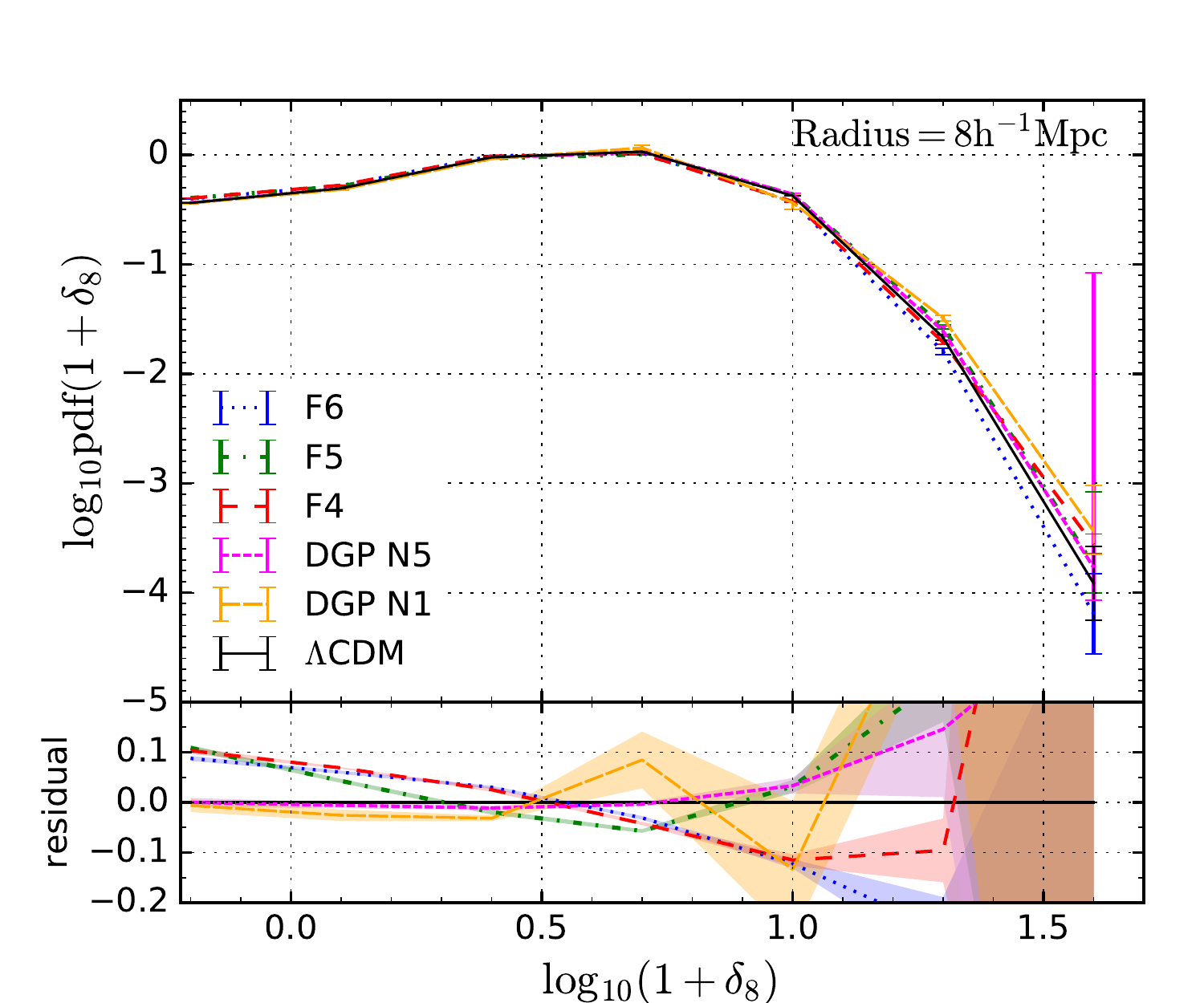}   
        \includegraphics[width=0.5\textwidth,angle=0]{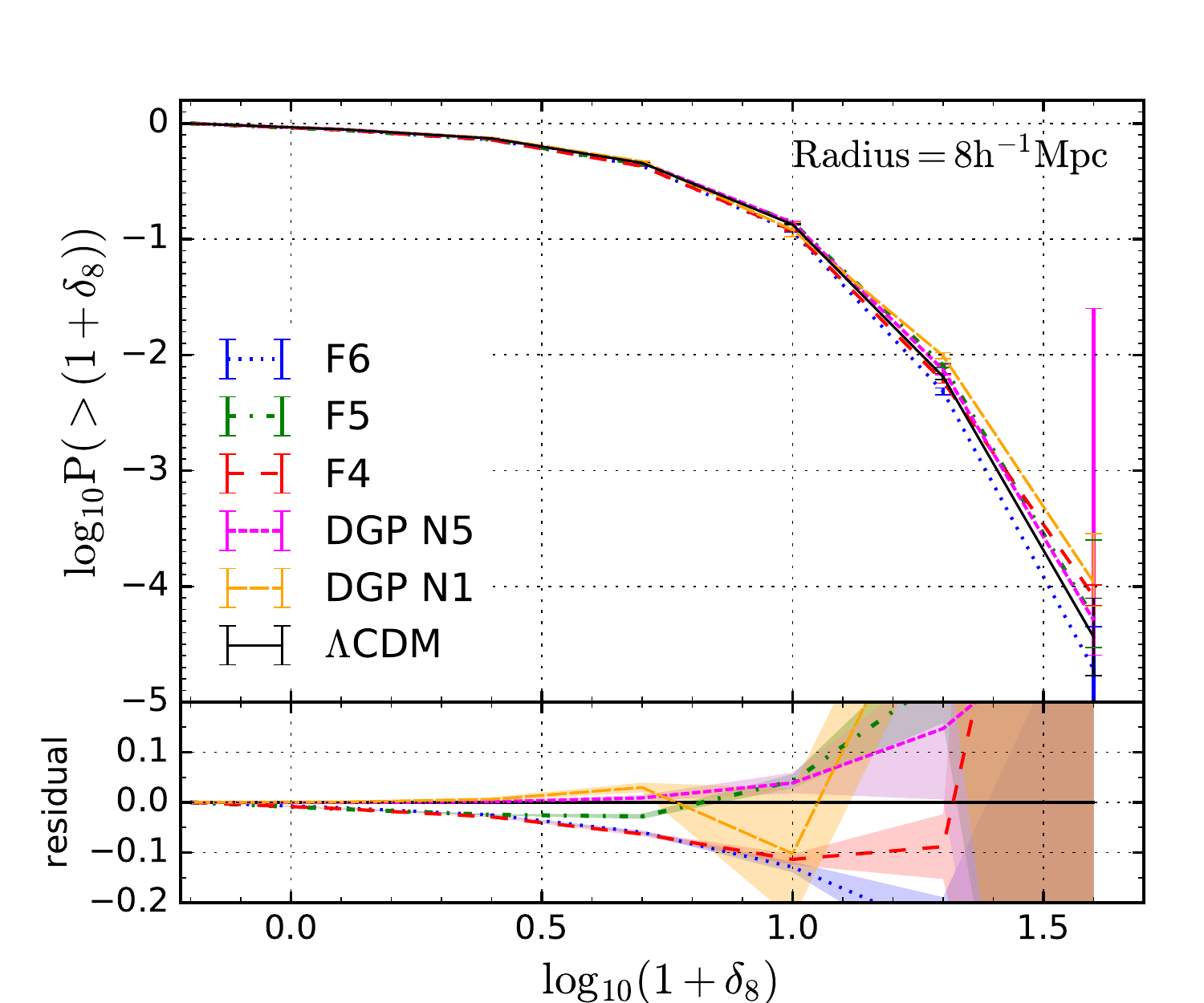} \\

     \end{tabular}

\caption{{\it Left panel:}  
Differential $(1+\delta_{8})$ probability functions of all type of galaxies for all the models, where overdensity $\delta_{8}$ is calculated  
within the sphere of radius $R =8 {\rm Mpc h^{-1}}$. {\it Right panel}: Cumulative $(1+\delta_{8})$ probability function. The binning is done with $0.3$ dex. 
}
\label{fig:pdf}
\end{figure*}

\begin{figure*}
     \centering
     \begin{tabular}{cc}
        \includegraphics[width=1\textwidth,angle=0]{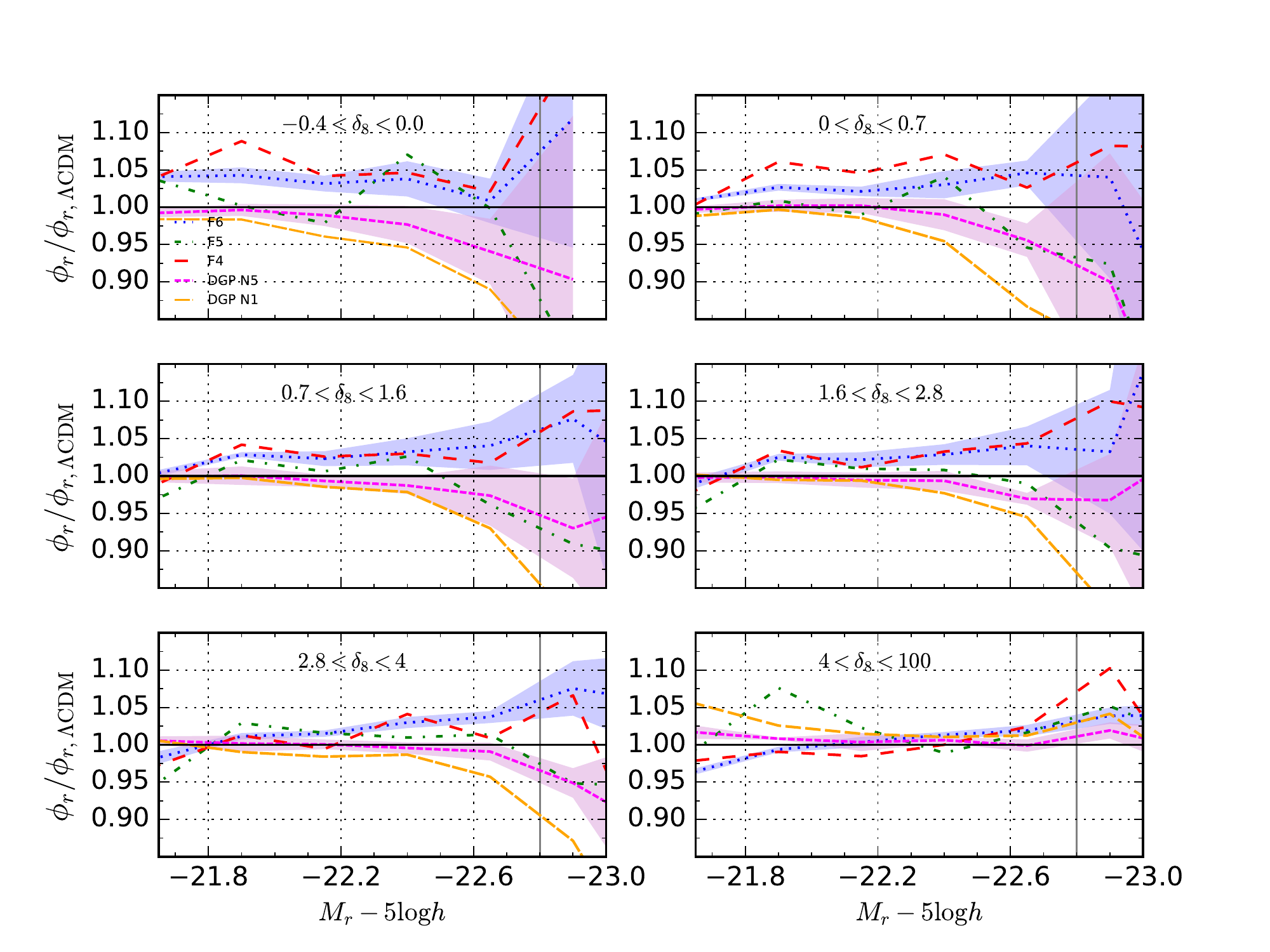}  

     \end{tabular}

\caption{GLFs in the $r$-band of each one of the MG models considered here with respect to the $\Lambda$CDM GLF in six different density environments determined within the sphere of $R= 8$ Mpc $h^{-1}$. The overdensities $\delta_8$ decrease from upper left to the right lower corners. The resulting GLFs are the mean of all the available simulation realizations.  The shaded regions show the respective error propagation from the standard deviation calculated from the realizations available for the F6 and N5 MG models, which are the closest in nature to the $\Lambda$CDM from the theoretical background. The error propagation for the other models is not shown to avoid overcrowding in the figures. The solid vertical line shows the threshold limit below which we cannot trust on the results due to uncertainties, coming from various factors, e.g SHAM technique, estimation of the $f_{sat}$ function, etc.}
\label{fig:GLFs1}
\end{figure*}

In this Section we present the resulting density distributions ${\rm pdf}(1+\delta_8)$ and the dependences of the
luminosity functions on environment for all the MG models
studied in this paper. We investigate whether the effects
of environment from the MG models lead to differences on the luminosity functions, as different gravity models posse different screening mechanisms that should have some signatures under different environmental conditions. 

\subsection{The Density Distribution}

We begin by describing the probability distribution of densities as measured for all the simulations.
The left panel of Figure \ref{fig:pdf} shows the probability distribution of galaxy densities, ${\rm pdf}(1+\delta_8)$ . Recall that densities were defined as the number of DDP neighbors within spheres of $R= 8 h^{-1}$ Mpc. The bottom
panel shows the difference with respect to the $\Lambda$CDM model. In general,
Figure \ref{fig:pdf} shows that the probability distribution for all the models are similar, though in more detail, some differences appear.   

Figure \ref{fig:pdf} shows that the largest significant differences between 
models appear for the low values of the overdensity, that is, for the void-like environments, $\delta_8<0$. 

Moreover, there is a systematic trend among the models; F4, F5 and F6 predict up to $10\%$ for
more low overdensities than the $\Lambda$CDM model, in contrast, the nDGP N1 and N5 models are slightly below the
$\Lambda$CDM model but barely indistinguishable. This is in agreement with what we expect; for $f(R)$ models in the  low dense environments screening mechanisms are not that much effective \citep[see e.g.,][]{Shi:2017pyd,Falck:2015rsa}. For  values of $\delta_8\sim 1$, F4 predicts an excess of low overdensities compared to F5, F6 and the $\Lambda$CDM. Thus, the differences observed above clearly show the dependence of 
the chameleon mechanism with the parameter $f_{R0}$ as well as with environment. 

While in case of nDGP models, the Vainshtein mechanism seems to work efficiently, we observe 
no environmental effect which is consistent with the previous results of \citet{Falck:2015rsa}. 
In denser environments, $\delta_8>4$, the above situation reverses and
F4 and F6 models predict a lower fraction of cluster-like environments, by $\sim10\%$, while F5  
predicts a larger number of high density environments, by $\sim20\%$. Note that
the excess of dark matter halos observed in Figure \ref{fig:Vmax} are within the 
magnitude range definition of our DDP population. While, at this point, it is not clear how
they would affect the definition of environment, one could argue that there is a 
non-negligible chance that the difference
observed above are simply the result of including those halos in our DDP definition. We will discuss further this point in Section
\ref{secc:discusion}.
Finally, as for 
nDGP N1, this model makes excursions around the $\Lambda$CDM expectations but at the largest 
density environments it predicts more structures with high $\delta_8$, $>20\%$. But
for the nDGP N5 we observe that is
consistent with the $\Lambda$CDM at all density environments and it is difficult to conclude whether there
is some environmental dependence for these models due the large errors in our determinations. 

The right panel of Figure \ref{fig:pdf} shows the cumulative probabilities $P(>\delta_{8}+1)$, while
the bottom panels present the residuals with respect to the 
$\Lambda$CDM model. While the trends observed in this figure are well understood from the
the density probability distribution (left panel), Poissonian errors have a lower impact
in $P(>\delta_{8}+1)$, which could be used for constraining gravity models. 

\subsection{The dependence of the luminosity function on environment in modified gravity models}
\label{secc:GLF_dependence_envirom}

As discussed above, we find evidence that the resulting galaxy 
density fields are different for different gravity models as well as for their
screening mechanisms. 
Next, we study the effect of the gravity models on the dependence of the luminosity function with environment.

The main goal of this work is to quantify the differences on the dependence of the $r$-band GLF with environment for all the gravity models studied here. Recall that Figure~\ref{fig:lum_LCDM} shows the GLFs under different environments ($\delta_8$ values) for the standard $\Lambda$CDM model.

The resulting differences of the respective GLFs in the different gravity models w. r. t. the $\Lambda$CDM model
are shown in Fig.~\ref{fig:GLFs1}. Each panel corresponds to the different 
environments, ranging from void-like structures, $-0.4<\delta_8<0$, to clusters-like
environments, $\delta_8>4$. The shaded regions show the $1 \sigma$ 
standard deviation from the 5 realizations for the F6 and N5 gravity models. 
We only present uncertainties in F6 and N5 to avoid overcrowding as they are expected to be the closer in nature to the $\Lambda$CDM model. 
We note, however, that we find similar uncertainties for all
the other gravity models. The solid vertical line in each panel shows the 
threshold limit, $M_{r} < -22.8$, below which we cannot trust the results due to sampling variance in the simulations.
In Appendix \ref{app:errors}, we show that when recovering the cumulative luminosity function from the simulations,
uncertainties due to sampling variance, $\sim5-10\%$, 
become relevant for bright galaxies.
Thus, as a conservative limit we use the threshold limit of $M_{r} < -22.8$,
as shown in Figure \ref{fig:cum_lum}.

In Figure~\ref{fig:GLFs1}, we observe that overall all MG models considered here
deviate from the standard $\Lambda$CDM model at {\it all density bins}. This
is especially true at low density environments where we find differences
of the order of $\sim10\%$ in some of the models, where the screening mechanisms are not efficient. At the high density environments, where the screening mechanisms are expected to be more efficient, the differences are around $\sim5\%$. 

In general the F4 and F6 models predict a higher number of galaxies for all magnitudes and for most of the environments. This situation is inverted at the highest density bin at which they predict a lower amplitude
for low luminosity galaxies but approaches to the $\Lambda$CDM model at the bright end. The model F5 is closer
to the $\Lambda$CDM model within $0\lesssim\delta_8\lesssim4$. At the lowest overdensity bin it seems to be higher 
than the $\Lambda$CDM model but this depends on luminosity, galaxies with $M_r\sim-22.5$ have
a maximum deviation of $\sim9\%$ from the $\Lambda$CDM model. At the
largest overdensity bin, $\delta_8>4$, F5 predicts $\sim9\%$ more 
galaxies at $M_{r}\sim-21.9$ but it is similar to the $\Lambda$CDM model
for brighter galaxies. 
As for nDGPs models, both N5 and N1, are almost indistinguishable to the 
$\Lambda$CDM model for magnitudes below $M_r\sim-22.2$. Nonetheless,
we find some deviations for brighter galaxies with differences around 
$8-10\%$ and $10-15\%$ respectively for N5 and N1 at a
fixed luminosity $M_{r}\sim -22.7$.

The above trends are surprising and unexpected, 
specially for the F4, F5 and F6 models.  
In fact, the F4 model is expected to deviate significantly from the $\Lambda$CDM, followed by F5 and F6 due to the chameleon screening mechanism. 
Note, however, that in Figure 
\ref{fig:Vmax} (see also Figure \ref{fig:HMF}) we showed that the differences
of the F4, F5 and F6 models w.r.t. $\Lambda$CDM depend on the
halo velocity (mass). Moreover we showed that for halos with 
$V_{\rm max}\sim 630$ km s$^{-1}$ (corresponding to $M_r\sim-22.2$
in terms of galaxies) there is a maximum excess of $\sim50\%$ of halos 
for the F5 model. As before, we do not see the above features in 
Figure \ref{fig:GLFs1}. In the case of the nDGP models, N5 was expected to be closer
to the $\Lambda$CDM than N1, something that we notice. However, 
it is not clear why they are below the predictions from the $\Lambda$CDM
model for brighter galaxies. In the next Section we discuss on the possible 
origin of the above tensions. 

Finally, despite of the details  
discussed above, Figure \ref{fig:GLFs1} shows that the differences in halo distributions among the different
gravity models are projected somehow in the dependence of the GLFs on environment
in such a way that is possible to distinguish between the different models by means
of observations. 

As noted previously, the most significant differences of modified 
gravity models w.r.t. the $\Lambda$CDM standard model is at low density 
environments, $\delta_8<1$. While there are hints of some differences
at high overdensity bins, most of the models are very close to the $\Lambda$CDM.
This is actually expected due to the enhancement of the screening mechanism in the 
different models that allows to recover GR in high-density regions.

The above trend with environment gives an idea on how does the MG 
affects the GLFs at different environments,  providing thus 
a valuable tool for distinguishing MG effects 
from $\Lambda$CDM. Moreover, based on our results, we propose that the dependence 
of the GLF on environment can be used to constrain screening mechanisms
along with the gravity models. 

\begin{figure*}
     \centering
     \begin{tabular}{cc}
        \includegraphics[width=1\textwidth,angle=0]{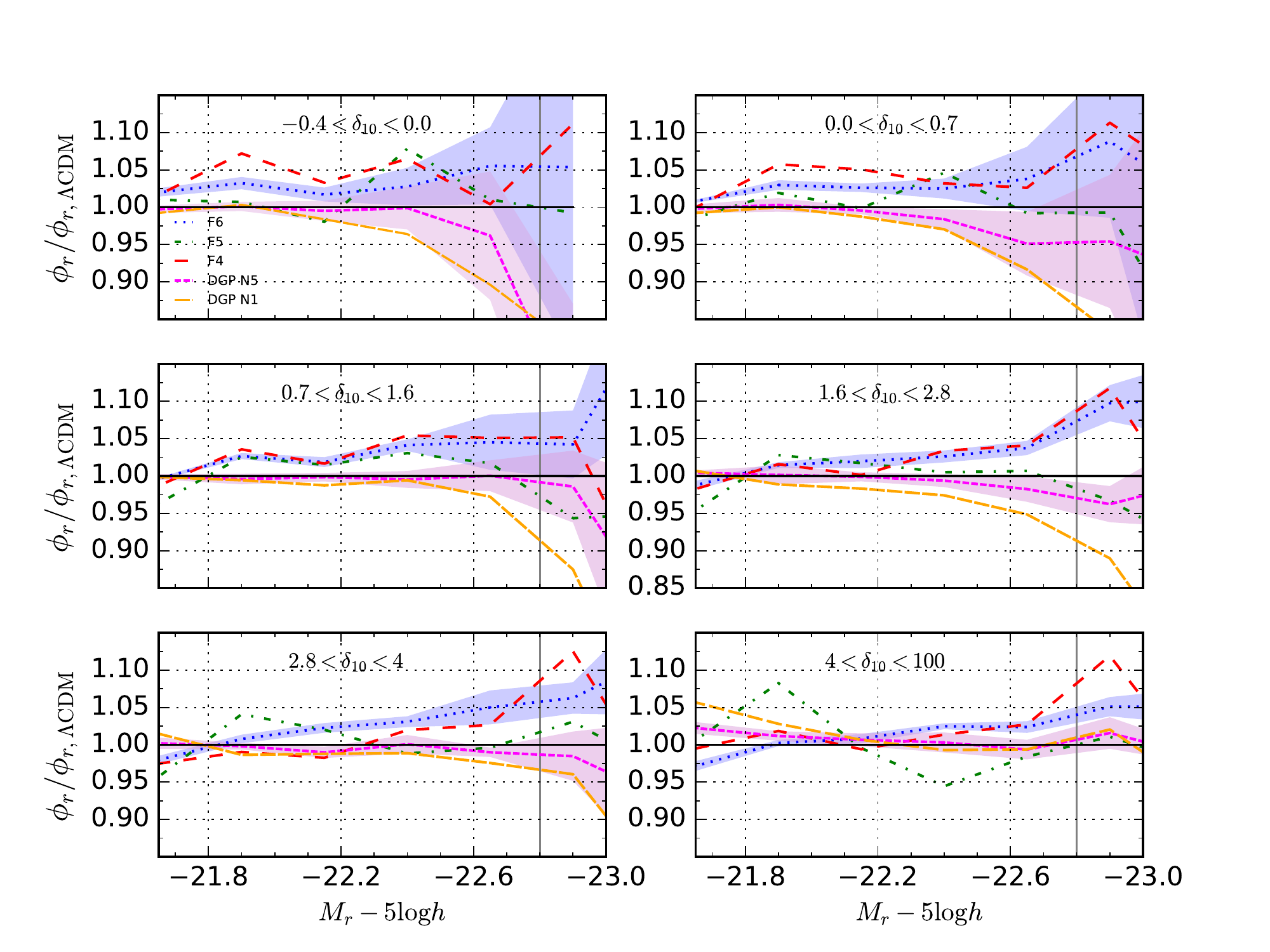}  
     \end{tabular}

\caption{Same as figure (\ref{fig:GLFs1}) but for a sphere of R= 10 Mpc $h^{-1}$. By varying from 8 to 10 Mpc $h^{-1}$, the radius of the sphere where the overdensities are measured, we do not see major differences for the MG models throughout all the different environments. 
}
\label{fig:GLFs2}
\end{figure*}

Similarly to Figure \ref{fig:GLFs1}, Figure \ref{fig:GLFs2} shows the 
dependence of the luminosity function with environment w.r.t. $\Lambda$CDM
but this time using an aperture sphere of $R= 10$ Mpc $h^{-1}$. Note that using larger apertures would tend to oversample voids
and thus increasing the effects of the fifth force from 
the MG models. 
Nonetheless, we observe
approximately the same trends as in the case of spheres of  $R= 8$ Mpc $h^{-1}$. 
Thus increasing the aperture radius do not significantly change our conclusions. 

\begin{figure*}
     \centering
     \begin{tabular}{cc}
       \includegraphics[width=0.9\textwidth]{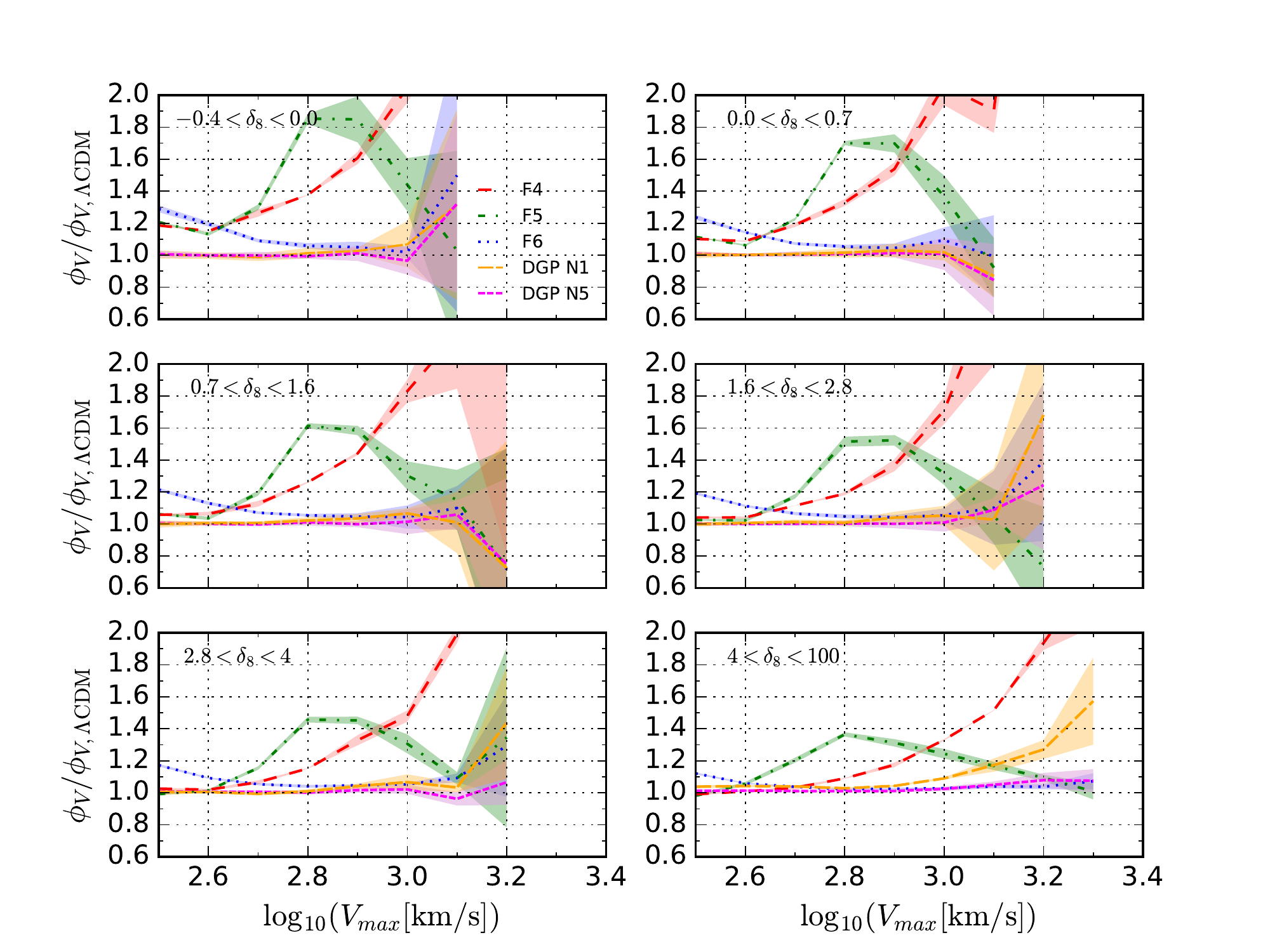}      \end{tabular}

\caption{Variations with environment the halo $V_{\rm max}$ functions of the MG models with respect to the $\Lambda$CDM. The shaded areas represent the error propagated from the standard deviations of all the  available realizations.  Some of the trends seen in Figure \ref{fig:Vmax} for halos in all the environments are enhanced in the low-density environments. While in high-density environments 
the fifth force is expected to be suppressed but we observe differences between
screening models and similar features to Figure \ref{fig:Vmax}.
}
\label{fig:vmax_enviroment}
\end{figure*} 
\begin{figure*}
     \centering
     \begin{tabular}{cc}
       \includegraphics[width=0.9\textwidth]{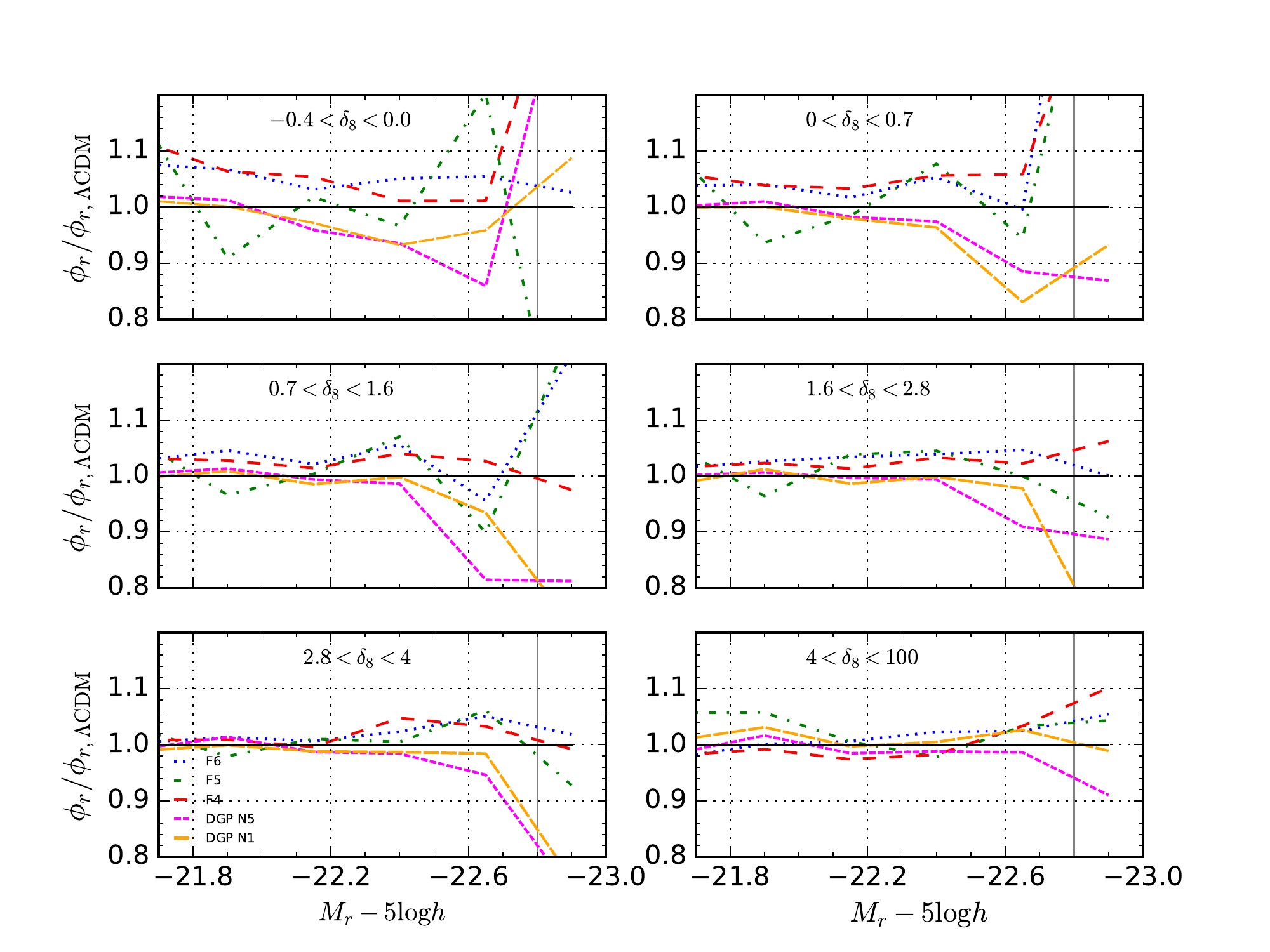}      \end{tabular}

\caption{As Fig. \ref{fig:GLFs1} but for the GLFs calculated from the respective halo $V_{max}$ functions as a function of environment, $\phi_V(V_{\rm max}|\delta_8$), and the $M_r-V_{\rm max}$ relation (see Eq. \ref{eq:SHAM_dif}). Only one realization for each models has been used for this plot. The results in this plot are similar to those obtained directly from the mock catalogs, shown in Fig. \ref{fig:GLFs1}.
}
\label{fig:test_magnitudes}
\end{figure*} 

\section{Discussion} 
\label{secc:discusion}

In the non-linear regime of structure formation, different screening mechanisms
predict that dark matter particles will cluster differently. In consequence,
as it has been shown by previous studies \citep[see e.g.,][]{Falck:2015rsa}, halos themselves cluster differently as well. 
In this paper, we study the environmental dependence of the GLFs as predicted by different gravity models.
Here we use two different
class of MG models and their screening mechanism, the $f(R)$ \citep{HuSawicky2007} and the nDGP 
\citep{Dvali2000}, in order to build mock galaxy catalogs and their corresponding galaxy density field. 

In the preceding Section, we conclude that the dependence of the 
GLF on environment is a valuable tool for constraining the screening
mechanisms in addition of the MG models. Nonetheless, we noted that 
the trends in the GLFs were counter-intuitive to what theoretically is expected, 
see Figure \ref{fig:Vmax} from Section \ref{sec:N-body}, 
specially for the F4, F5 and F6 models. There, we discussed that 
the chameleon screen mechanism tends to produce an excess of abundance of dark matter
halos around some characteristic velocity/mass that depends on the present day value of the scalaron field, \citep{Li_Efstathiou_2012}. In the halo velocity/mass range 
of the suite of $N-$body simulations we are employing for this paper, we observe 
that the maximum excess of dark matter halos for the F5 model is of $\sim50\%$, w.r.t. 
$\Lambda$CDM, at $V_{\rm max}\sim 600$ km s$^{-1}$, which corresponds to halos with 
$M_{\rm 200c}\sim 6\times 10^{13}M_{\odot}$; see Figures \ref{fig:Vmax} and
\ref{fig:HMF}, respectively. On the other hand, F6 and F4 models show an excess of dark 
matter halos respectively of $\sim20\%$ (at low velocities/masses) and more
than $100\%$ (at high velocities/masses). In the case of the nDGP models, N1 predicts an
excess of halos at the high-mass end while N5 is almost indistinguishable from
the $\Lambda$CDM. Further, we discuss why
we do not recover the above trends into the predicted GLFs. 

Figure \ref{fig:vmax_enviroment} presents the dependence of the 
halo velocity functions for the various MG models w.r.t. $\Lambda$CDM on the environment. 
Note that we employed the same definition of environments, $\delta_8$, 
derived for their host galaxies from the
previous Section. The halo velocity functions were derived 
using the same methodology as described in Section \ref{Method_GLFs_enviroment}.
Figure \ref{fig:vmax_enviroment} shows that the trends observed 
in Figure \ref{fig:Vmax} are replicated in all the environments. As expected, at the low-density environments the effects of
the fifth force are more relevant leading to a larger differences w.r.t. $\Lambda$CDM than at the high-density environments, where
the fifth force is suppressed and the screening mechanisms are more efficient. Note, however, that even in the highest overdensity bin we do observe similarities with 
Figure \ref{fig:Vmax}. Naively, the above signatures are what we would expect to be printed in the dependence of the GLFs with environment. 
Therefore, we discard that the methodology employed in Section
\ref{Method_GLFs_enviroment} is responsible for erasing 
the features observed in Figure \ref{fig:Vmax}. Moreover, we also discard that choosing
our DDP population within the magnitude range at which
the F5 model shows an excess of halos, 
is not introducing an extra source of bias between the F5 and
the other models.
Next, we investigate whether the differences on the 
$V_{\rm max}-M_r$ relations are responsible for the counterintuitive results
of the dependence of the GLFs with environment.  

In this paper we derived the $V_{\rm max}-M_r$ relationship via SHAM under the assumptions of
zero scatter and {\it separately} for all the gravity models.
Thus, in this approach, halos with identical $V_{\rm max}$ 
will host galaxies with identical luminosities $M_r$, no matter what their environmental density is. In other words, the
$V_{\rm max}-M_r$ relation is independent of environment. 
Assuming that we have a model for the dependence of the velocity function with
environment $\phi_{V}\left(V_{\rm max}|\delta_8 \right)$ and using the universality of the 
$V_{\rm max}-M_r$ relation, we can thus derive
the dependence of the GLF with environment as:
\begin{equation}
\phi_{r}(M_r|\delta_8) = \phi_{V}\left(V_{\rm max}(M_r)|\delta_8 \right) \left| \frac{d\log V_{\rm max}(M_r)}{dM_r} \right|.
\label{eq:SHAM_dif}
\end{equation}
Notice that the above equation is simply the differential form of SHAM. 
When examining 
Equation (\ref{eq:SHAM_dif}), it clearly shows that when two models
have different $V_{\rm max}-M_r$ relationships, the observed features
from these models in their corresponding velocity functions will not be directly
projected into their GLFs. This is due to the non-trivial 
relationship between galaxies (magnitudes) and halos (velocities), depending
strongly not only in the functional forms of 
the $V_{\rm max}-M_r$ relations but also on their slopes. 

To understand the above, 
imagine that we want to compare two models with 
different $V_{\rm max}-M_r$ relations
such that Model 1 has a higher amplitude than Model 2.
That is, at a fixed $V_{\rm max}$ the Model 1 host
brighter galaxies than the Model 2. Equivalently, at a fixed 
luminosity we find that $V_{\rm max,1}<V_{\rm max,2}$, where the 
subscripts '1' and '2' indicate the models. Equation 
(\ref{eq:SHAM_dif}) shows that when comparing two models 
with identical magnitudes (this is actually our 
situation in Figure \ref{fig:GLFs1} where we are comparing 
different models at a fixed magnitude) 
the differences observed in Figures \ref{fig:Vmax} and \ref{fig:vmax_enviroment}, 
at a fixed $V_{\rm max}$, will project differently into the GLFs,
due to the shift in the halo velocities by simply fixing the galaxy magnitudes. 
This could explain the observed trends from Figure \ref{fig:GLFs1}. 
The mapping between $V_{\rm max}$ and $M_r$ would perhaps in some cases compress, stretch, squash or just shift, 
or a combination of all of them, the features observed in Figures \ref{fig:Vmax} and \ref{fig:vmax_enviroment}.  

We test the above idea, by choosing one (from the five realizations) of the simulations 
for each gravity models and by employing 
Equation (\ref{eq:SHAM_dif}) combined with our determinations
of $\phi_{V}\left(V_{\rm max}|\delta_8 \right)$, similarly to Figure 
\ref{fig:vmax_enviroment}, and the $V_{\rm max}-M_r$ relations for all the 
gravity models, see Figure \ref{fig:Magr}. The resulting GLFs as a function of
environment, $\phi_{r}(M_r|\delta_8)$, are shown in Figure \ref{fig:DGLS_enviroment}
from Appendix \ref{app:luminosity_Vmax}. Note that our results based on SHAM, 
Equation (\ref{eq:SHAM_dif}), are very similar to the direct measurements from the simulations. 
Next we study the ration w.r.t. $\Lambda$CDM. 

Similarly to Figure \ref{fig:GLFs1}, 
Figure \ref{fig:test_magnitudes} shows the dependence of the GLF with environment from 
Equation (\ref{eq:SHAM_dif}) w.r.t. $\Lambda$CDM. Observe that we are replicating the
same features in both figures. The above confirms that the differences in the 
$V_{\rm max}-M_r$ relations of the gravity models are responsible for the intriguing
and counter-intuitive results from Section \ref{secc:GLF_dependence_envirom}. The differences within $10\%$ among the models are being observed in the GLFs under different environments with the fact that different gravity models lead to have different Mr-Vmax relationships.

We end this Section by emphasizing that the features derived for the distributions of dark 
matter halos would not necessarily map directly into their host galaxies. As we
have discussed, the mapping between galaxies and dark matter halos is not trivial and it depends
on the gravity model. In order to compare predictions from different gravity models,
one should use the correct galaxy-halo connection for each model. Otherwise, one would be pruned to 
draw wrong conclusions on the real viability of one model over the others. 
That would be the case when using HOD parameters derived from the $\Lambda$CDM model
but employed in other MG models, see also the discussion in Section \ref{sec:HOD_parameters}. 

\section{Summary and Conclusions} 
\label{sec:summary_and_discussion}

We have studied the differences of several halo and galaxy distributions predicted within the
context of two classes of MG models and their respective screening mechanisms. We explored (i) the $f(R)$ model of \cite{HuSawicky2007} with $n=1$ and three different $|f_{R0}|$ values: F6$=10^{-6}$, F5$=10^{-5}$, and F4$=10^{-4}$ (see Eq.\ref{HSmodel}), and (ii) the normal branch of \cite{Dvali2000} (nDGP) Braneworld model, with $r_{c} H_{0}/c = 5$ and $1.0$, denoted respectively by N5 and N1. We used a large suite of high-resolution $N-$body cosmological simulations for these MG models and for the standard $\Lambda$CDM model. The simulations were presented in \citet{Li2012,Li2013}. Each model has five different realizations, which are ran using slightly different random phases for the initial conditions, except for F4 model that has only two
realizations. We used these realizations to determine uncertainties from sampling variance. 

The dark matter (sub)halos 
were populated with galaxies by means of the Subhalo Abundance Matching (SHAM). For the SHAM, we used the halo maximum circular 
velocity ($V_{\rm max}$) function from the simulations and the $r-$band GLF from SDSS. As the result, and following \cite{Rodriguez2012},
we obtained the $M_r-V_{\rm max}$ relationships for both the central and satellite galaxies (halos and subhalos). These relationships connect galaxies with (sub)halos, in such a way that the spatial clustering of galaxies at different $r$-band luminosities can be measured in the simulations. For all the gravity models studied here, the predicted projected two-point correlation functions (figure \ref{fig:Projected2points}) agree with observational determinations, with differences at some scales up to $\sim 30\%$, and with the deviations from them being similar for all the models. This shows that the employed SHAM method is relatively robust and that the systematic differences with observations do not introduce biases in our further comparative analysis among the different gravity models. 

Further, we characterized the galaxies in the simulations by their environmental density, $\delta_8$, by counting 
neighbours in spheres of $R = 8$ Mpc $h^{-1}$, and calculated the GLFs as a function of environment with the aim of exploring whether the dependence of the GLF on environment changes among the different gravity models, providing possible observational signatures to constrain them. Our main results and conclusions are as  follows:

\begin{itemize}
\item The $V_{\rm max}$ function, $\phi_V(V_{\rm max})$, and the halo mass function, $\phi_h(M_{\rm 200c})$, for halos and subhalos depend on the gravity models.
The F4 model predicts $\sim50\%$ more halos and subhalos than the $\Lambda$CDM model
around $V_{\rm max} \sim 1000~{\rm km~s^{-1}}$, the difference increasing for larger values of $V_{\rm max}$.
For the F5 model, there is an excess of halos/sub-halos w.r.t. the $\Lambda$CDM model up to $50\%$ at the $V_{\rm max}$ range of
$400-1000~{\rm km~s^{-1}}$. 
We observe qualitatively similar differences in the halo mass function but with smaller amplitude.

This is because the halos in these $f(R)$ models are also more concentrated (higher $V_{\rm max}$ values for a given halo mass), so that besides the abundance excess, they have larger velocities than in the $\Lambda$CDM model.  
 
These differences are expected due to the inefficiency of the screening mechanism for these models in the relevant 
$V_{\rm max}$/mass regime. On the other hand, the F6 model predictions are similar to those of the $\Lambda$CDM model due to the strong screening mechanism acting on the $V_{\rm max}$/mass regime explored in the simulations. 
Finally, the N1 and N5 models remain indistinguishable from 
the $\Lambda$CDM model throughout most of the $V_{\rm max}$/mass ranges due to the effective
suppression of the fifth force under the Vainshtein screening method employed by these models.

\item The obtained $M_{r}-V_{\rm max}$ relationships of central and satellite galaxies vary among the various 
gravity models. The differences are not larger than $\sim1\%$, which
correspond to differences of $\sim0.2$ mags in  $M_{r}$. These results are mainly
consequence of the differences among the maximum circular velocity functions reported above. 

\item The variation of the $M_{r}-V_{\rm max}$ relationships with gravity
models affects the results on the halo occupation distributions, HOD. 
The maximum variation in   
the best fit values of the HOD parameters among the different gravity models are around $\sim40\%$ for $M_{1}$;
$\sim 35\%$ for $M_{\rm cut}$; $\sim 26\%$ for $M_{\rm min}$; 
$10\%$ for $\alpha$; and $20\%$ for $\sigma_{\log M}$.
We observe differences in the halo occupation numbers between the $f(R)$ models and the $\Lambda$CDM one up to 50\%. The major differences (a defficit of centrals) are at $ M_{200c}< 10^{13} M_{\odot}h^{-1}$, and also at the high-mass end for the F4 model. For the nDGP models the differences are below the $10\%$. We stress that since the HOD numbers depend on the MG, it is not
correct then to apply $\Lambda$CDM-based HOD parameters for seeding galaxies in halos from simulations for the MG models. 

\item The galaxy overdensity distribution, ${\rm pdf}(1+\delta_8)$, varies with the gravity model.
For the F4, F5 and F6 models, there is a $10\%$ excess of void-like regions compared to the $\Lambda$CDM.
In contrast, at the high-end of the $(1+\delta_8)$ distribution, for F4 and F6, 
there is a $\sim10\%$ deficit of high-density environments but for F5 there is a $\sim20\%$ excess of
high-density environments. For the DGP N1 and N5 models, the overdensity distributions are
similar to those predicted for $\Lambda$CDM, excepting at the high-density end, where the 
variance is large. 

\item The various gravity models analyzed here predict a different dependence 
of the GLF with environment. The most significant difference w.r.t. the $\Lambda$CDM is at the 
low density environments, $\delta_8\lesssim 1$. The F4 model has an excess of $\sim 5-10\%$ at the lowest density bins and at all luminosities, followed by the F6 model. Contrary to the naive expectation, F5 is closer to the $\Lambda$CDM prediction than F6 along with some fluctuation in F5.
The DGP N5 and N1 models are closer to the $\Lambda$CDM model for lower-luminosity galaxies but we observe a deficit of
high-luminosity galaxies in almost all the environments, $\sim9\%$ and $\sim12\%$ 
for N1 and N5 models, respectively. 

\item We have discussed the counterintuitive results on the expected
dependence of the GLFs with environment and screening mechanisms, 
specially for the sequence of models F6, F5, and F4.
The dependence of the halo $V_{\rm max}$ function with environment, $\phi_V(V_{\rm max}|\delta_8)$, is mapped into a dependence of the GLFs by environment, $\phi_r(M_r|\delta_8)$, through the $V_{\rm max}-M_r$ relationship (Eq. \ref{eq:SHAM_dif}) in a very non linear way because the latter has a non-trivial shape that varies among the different gravity models and their screening mechanisms. The effects of the shape and differences in the $V_{\rm max}-M_r$ relations could result in some cases in the compression, stretch, squash or just shift (or a combination of all of them) of the features produced by the screening mechanisms in the $\phi_V(V_{\rm max}|\delta_8)$ function.
 
\end{itemize}

Our results are in general consistent with the findings of \citet{Hernandez-Aguayo:2018yrp}, where the authors studied the impact of $f(R)$ gravity on galaxy clustering using marked correlation functions. In the marked correlations approach, one gives a `weight' or `mark' to each galaxy as a function of the environment \citep{Sheth:2005aj}, with the idea to up-weight 
low-density regions to boost the MG signal in galaxy clustering \citep{White:2016yhs}. \citet{Hernandez-Aguayo:2018yrp} found that up-weighting low- and intermediate-density regions it is possible to find measurable differences between $f(R)$ gravity models and GR ($\Lambda$CDM). This result is similar to the differences found here in the GLFs as a function of $\delta_8$ and $\delta_{10}$ (see Figs.~\ref{fig:GLFs1} and \ref{fig:GLFs2}), especially in the bins: $-0.4<\delta_{8/10}<0$, $0<\delta_{8/10}<0.7$ and $0.7<\delta_{8/10}<1.6$.

Our simple approach of analyzing the GLFs under different environment can provide a 
complimentary test to other existent non-linear observables such as galaxy
clustering redshift-space distortions \citep{Vlah:2018ygt,Hernandez-Aguayo:2018oxg}, weak lensing measurements \citep{Shirasaki2015}, cluster abundances \citep{Cataneo:2014kaa} and the marked correlation functions \citep{White:2016yhs,Hernandez-Aguayo:2018yrp}. Indeed, our measurements quantify the effect of 
  MG models and their screening effect as a function of galaxy environment, 
  especially in the low (void like) and high (cluster like) density regions.  
Perhaps, combining various observables will improve the constraints on the different MG
models. In addition, the fact that the HOD parameters depend on the gravity model, as we have found,
observational results on the conditional luminosity/stellar mass functions \citep[e.g.,][]{Yang+2007} would be helpful for constraining the models. 

Finally, this paper is the first on a series for examining the effects of the screening mechanism of the MG models at the level of 
galaxy properties, particularly the dependence of the GLF with environment. While our results show that it is, in 
principle, possible to use the observed GLFs to establish limits or to constrain MG models, even though we have ignored errors from observations. We are currently developing realistic galaxy mocks
by including random errors from magnitude and redshift determinations in order to test the viability of using the above idea with current facilities such as the SDSS and DES or even with future surveys such as DESI. 
In addition, here we mainly focused on the $r$-band magnitude but the analysis can be easily extended to other bands
or even at the level of galaxy stellar mass. Considering the robustness of the SHAM method mentioned before, performing HOD directly on MG models is our next future plan.

\section*{Acknowledgements}
NCDevi thanks Hector Ibarra and A. R. Calette for useful comments and discussions. NCDevi acknowledges support from a DGAPA-UNAM post-doctoral fellowship and CONACyT Fronteras de la Ciencia grant 281. ARP and VAR
acknowledges support from UNAM PAPIIT grant IA104118 and from the CONACyT 'Ciencia Basica' grant 285721. OV acknowledges support from UNAM PAPIIT grant IN112518. CH-A acknowledges support from the Mexican National Council of Science and Technology (CONACyT) through grant No. 286513/438352. Some of the simulations used in this worked were run on the OKEANOS supercomputer hosted by the Interdisciplinary Centre for Mathematical and Computational Modelling, University of Warsaw, kindly made available to us by Wojciech Hellwing.  

\bibliographystyle{mn2e} 
\bibliography{refs}


\appendix

\section{Halo Mass Functions and halo concentrations from the gravity models}
\label{app:HMF_cvir}

\begin{figure*}
     \centering
     \begin{tabular}{cc}
\includegraphics[width=0.5\textwidth,angle=0]{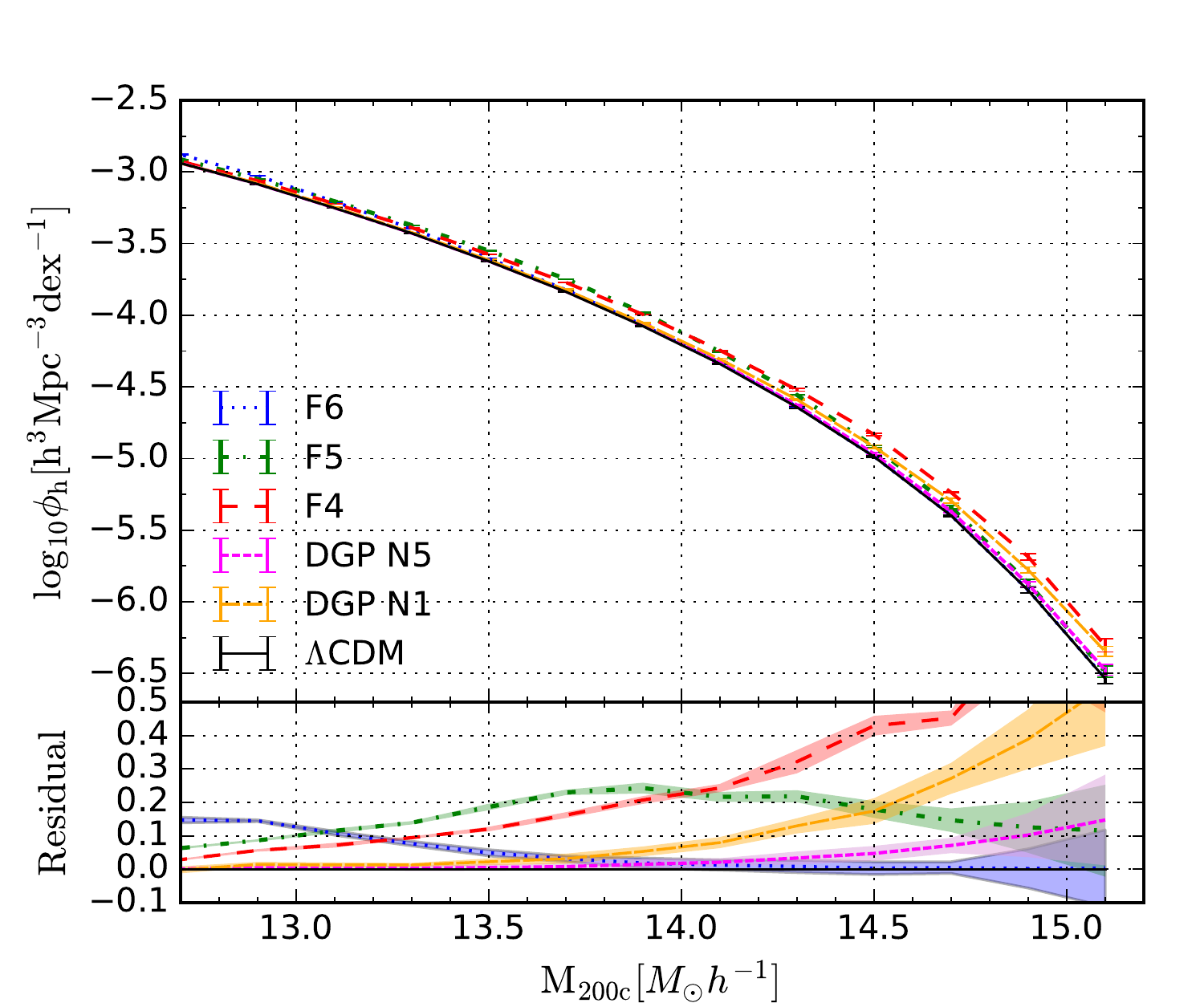}  

\includegraphics[width=0.5\textwidth,angle=0]{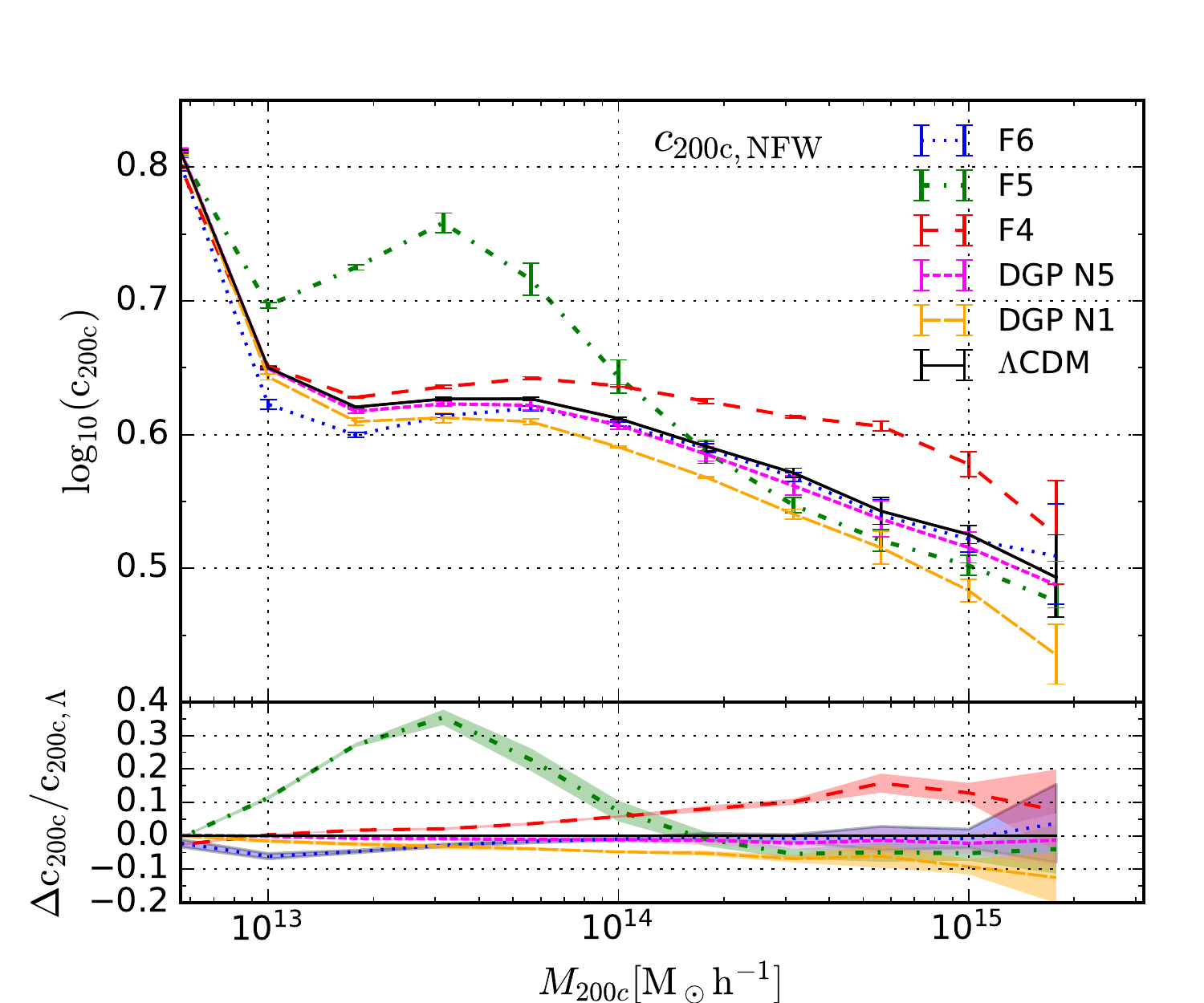} 
       
     \end{tabular}

\caption{{\it Left panel:}The differential halo mass $M_{200c}$ function of the dark matter halos for all the  MG models including $\Lambda$CDM at redshift, $z =0.0$. {\it Right panel}:
The dark matter halo concentrations as a function of $M_{200c}$ at redshift $z=0.0$
Their relative differences w.r.t the $\Lambda$CDM is shown in their lower panels with the shade regions representing the error propagated from 1-$\sigma$ standard deviations of all the realizations. For this particular simulation resolution, F5 model is showing the highest deviation $20-25\%$ from $\Lambda$CDM in the mass ranges $10^{13.5}M_{\odot}h^{-1}<M_{200c}<10^{14.5}M_{\odot}h^{-1}$ even larger deviation then F4, clearly showing the scales where the chameleon screening mechanism is inefficient for F5. F4 starts to deviate significantly from $10^{13.5}M_{\odot}h^{-1}<M_{200c}$ and continue to grow the deviation.}
\label{fig:HMF}
\end{figure*} 

The left panel of Figure \ref{fig:HMF} shows the halo mass function, $\phi_h(M_{200c})$, for 
all the gravity models employed in this paper. The bottom panel shows the relative differences with respect 
to the $\Lambda$CDM model. The trends noted for the halo velocity
functions, Figure \ref{fig:Vmax}, are similarly observed for the halo mass function, though the relative differences are 
of lower in amplitude for the latter. 
The right panel of Figure \ref{fig:HMF} shows the mean halo concentration--mass relation measured from the 
simulations corresponding to all the gravity models studied here. In the bottom panel, the relative differences
with respect to the $\Lambda$CDM model are shown. Note that the halo concentrations for the F5 model are significantly higher for masses $\sim 10^{13}-10^{14} M_{\odot}h^{-1}$, the same mass range where the halo mass function presents an excess. Similarly, for the F4 model,
the halos become more and more concentrated than in  $\Lambda$CDM for masses $\ga 3\times 10^{13} M_{\odot}h^{-1}$, the same mass range, where
the halo mass function deviates significantly from the one of the $\Lambda$CDM model. Therefore, the respective halos in these MG models are not only more abundant than in the $\Lambda$CDM model, but also more concentrated; the latter implies that for the same halo mass, $V_{\rm max}$ is higher in these MG models. As the result, the differences in the halo mass functions of the MG models with respect to the $\Lambda$CDM will further increase when plotting the halo $V_{\rm max}$ functions.  The trends seen in Figure \ref{fig:HMF} are mainly 
the result of the behaviour of the screening mechanisms, as discussed in Section \ref{halo-demographics}. 

\section{The Impact of poissonian Error}
\label{app:errors}


\begin{figure*}
     \centering
     \begin{tabular}{cc}
       \includegraphics[width=\columnwidth]{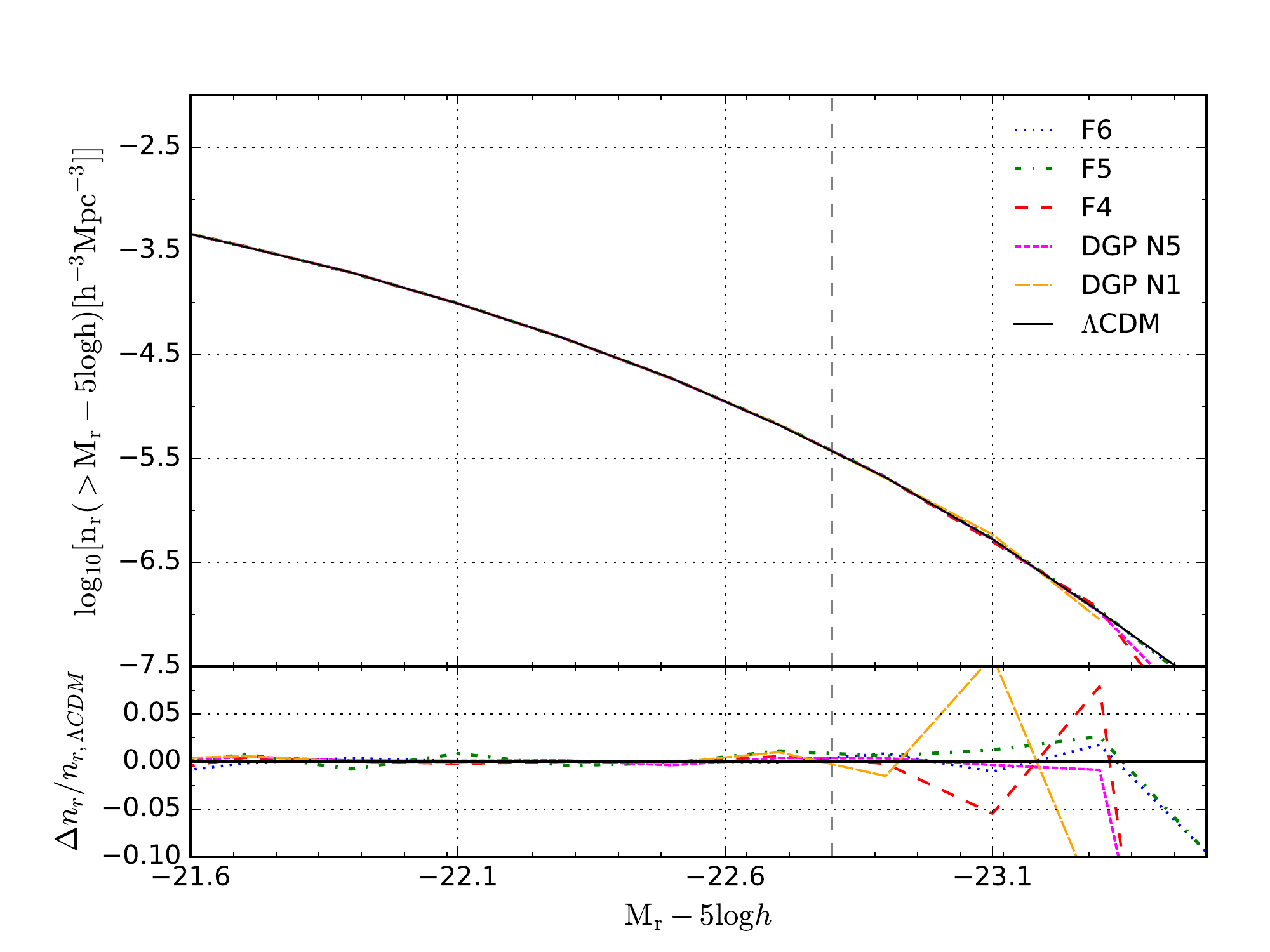}  
     \end{tabular}
\caption{The cumulative $M_{r}- 5\log h$ function of all type galaxies for all the models considered. The relative differences w. r. t. $\Lambda$CDM is shown in the lower panel. The dashed vertical line marks the threshold limit, below the threshold $M_{r} < -22.8$ the mock catalogs we generated using the SHAM technique can be biased by the Poisson noise of the cumulative function. This figure shows that the results we obtained within the ranges $-21.6< M_{r} -5\log h < -22.8$ can be trust to understand the intrinsic effect of MG in our analysis.}
\label{fig:cum_lum}
\end{figure*} 

The cumulative GLFs for all the gravity models are presented in figure(\ref{fig:cum_lum}) along with their relative differences w. r. t. $\Lambda$CDM in the lower panel. This figure shows how accurately the SHAM technique has been applied to generate the mock galaxies catalogs irrespective of MG. One can notice from the lower panel of the figure that the noise level remains within $1\%$ until $M_{r} \sim -22.8$; afterward the Poissonain error becomes prominent. Hence, we  mark this value as our threshold limit on $M_{r}$ beyond which our results can be biased by the Poissonian errors. Thus, the results we analysis below this limit $-22.8 > M_{r} -5\log h$ are marginal.

\section{Luminosity functions from the $V_{max}$ functions and $M_{r}-V_{max}$ relation}
\label{app:luminosity_Vmax}
\begin{figure*}
     \centering
     \begin{tabular}{cc}
       \includegraphics[width=0.9\textwidth]{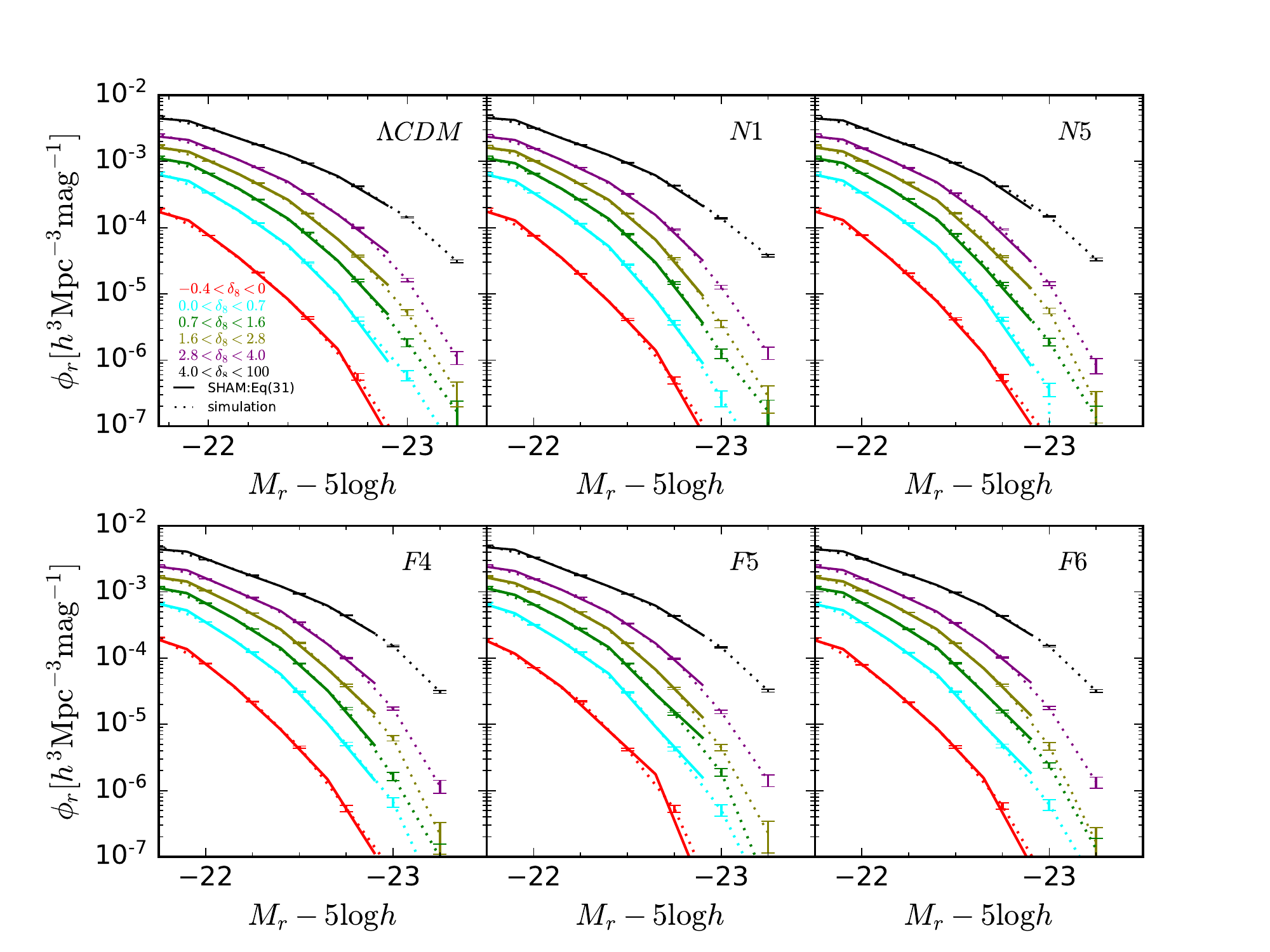}       \end{tabular}
\caption{Comparison of the GLFs of all the models calculated using eq.(\ref{eq:SHAM_dif}) (solid lines) with the one measured from our mock catalogs of galaxies (dotted lines) under six different density environments. The result is an outcome of one realization. Both are come out to consistent be with each other. }
\label{fig:DGLS_enviroment}
\end{figure*}

The GLFs of all models calculated using eq.(\ref{eq:SHAM_dif}) under different density environments for one realization of the simulations are shown in fig(\ref{fig:DGLS_enviroment}), presented as the solid lines in each panel. We also plot the corresponding GLFs measured from the simulation-based mock galaxy catalogs (dashed lines in the figures). As seen, the results are quite consistent with each other.

\section{Tables for the $M_r-V_{max}$ relationships separately for centrals and satellites}

The $M_r-V_{max}$ relationships for central and satellite obtained from SHAM method separately for all the gravity models 
are tabulated in Tables \ref{table:L-Vmax_central} and \ref{table:L-Vmax_satellite} respectively. 
We report the mean values for the realizations available for our suite of $N-$body simulations. Recall
that $r-$band magnitudes were derive for a redshift rest-frame of $z=0$. 

\begin{table*}
\centering
\begin{tabular}{c c c c c c c}
\hline
\hline
$V_{max}$ & $M_{r}-5\log_{10}h$(GR) & $M_{r}-5\log_{10}h$(F4) &$M_{r}-5\log_{10}h$(F5) &$M_{r}-5\log_{10}h$(F6)& $M_{r}-5\log_{10}h$(N1) & $M_{r}-5\log_{10}h$(N5) \\
\hline
\hline
$317.024$ & $-21.5304$ & $-21.5003$ & $-21.489$ & $-21.5003$ & $-21.5287$ & $-21.5278$\\ 
$339.697$ & $-21.5933$ & $-21.5624$ & $-21.549$ & $-21.5624$ & $-21.594$ & $-21.5906$\\ 
$363.992$ & $-21.6754$ & $-21.6423$ & $-21.6239$ & $-21.6423$ & $-21.6789$ & $-21.6725$\\ 
$390.024$ & $-21.7640$ & $-21.7284$ & $-21.7035$ & $-21.7284$ & $-21.7688$ & $-21.761$\\ 
$417.919$ & $-21.8494$ & $-21.8107$ & $-21.778$ & $-21.8107$ & $-21.8541$ & $-21.8462$\\ 
$447.808$ & $-21.9272$ & $-21.8857$ & $-21.846$ & $-21.8857$ & $-21.932$ & $-21.9239$\\ 
$479.835$ & $-22.0002$ & $-21.9556$ & $-21.910$ & $-21.9556$ & $-22.0051$ & $-21.9967$\\ 
$514.152$ & $-22.0714$ & $-22.0234$ & $-21.976$ & $-22.0234$ & $-22.0768$ & $-22.067$\\ 
$550.924$ & $-22.1422$ & $-22.0912$ & $-22.045$ & $-22.0912$ & $-22.148$ & $-22.138$\\ 
$590.326$ & $-22.213$ & $-22.159$ & $-22.118$ & $-22.158$ & $-22.219$ & $-22.209$\\ 
$632.546$ & $-22.284$ & $-22.226$ & $-22.195$ & $-22.226$ & $-22.2898$ & $-22.279$\\ 
$677.785$ & $-22.355$ & $-22.2937$ & $-22.275$ & $-22.298$ & $-22.3602$ & $-22.3496$\\ 
$726.259$ & $-22.425$ & $-22.3607$ & $-22.3571$ & $-22.3607$ & $-22.4304$ & $-22.4193$\\ 
$778.201$ & $-22.4962$ & $-22.4279$ & $-22.4387$ & $-22.4279$ & $-22.501$ & $-22.489$\\ 
$833.858$ & $-22.566$ & $-22.495$ & $-22.5189$ & $-22.4953$ & $-22.5705$ & $-22.5587$\\ 
$893.495$ & $-22.636$ & $-22.562$ & $-22.597$ & $-22.562$ & $-22.640$ & $-22.6288$\\ 
$957.397$ & $-22.7092$ & $-22.6318$ & $-22.676$ & $-22.632$ & $-22.712$ & $-22.701$\\ 
$1025.869$ & $-22.783$ & $-22.7047$ & $-22.757$ & $-22.705$ & $-22.786$ & $-22.774$\\ 
$1099.238$ & $-22.861$ & $-22.779$ & $-22.8398$ & $-22.779$ & $-22.863$ & $-22.851$\\ 
$1177.855$ & $-22.939$ & $-22.856$ & $-22.923$ & $-22.856$ & $-22.943$ & $-22.9287$\\ 
$1262.095$ & $-23.0144$ & $-22.929$ & $-23.001$ & $-22.929$ & $-23.0188$ & $-23.003$\\ 
$1352.359$ & $-23.088$ & $-22.9976$ & $-23.076$ & $-22.997$ & $-23.094$ & $-23.076$\\ 
$1449.079$ & $-23.1632$ & $-23.0706$ & $-23.151$ & $-23.0706$ & $-23.163$ & $-23.1504$\\ 
$1552.716$ & $-23.250$ & $-23.156$ & $-23.239$ & $-23.156$ & $-23.246$ & $-23.236$\\ 
$1663.765$ & $-23.355$ & $-23.266$ & $-23.347$ & $-23.266$ & $-23.355$ & $-23.3389$\\ 
$1782.756$ & $-23.467$ & $-23.384$ & $-23.464$ & $-23.384$ & $-23.474$ & $-23.448$\\ 
$1910.258$ & $-23.555$ & $-23.481$ & $-23.557$ & $-23.481$ & $-23.573$ & $-23.536$\\ 
$2046.878$ & $-23.598$ & $-23.526$ & $-23.602$ & $-23.526$ & $-23.619$ & $-23.578$\\ 
$2193.269$ & $-23.572$ & $-23.495$ & $-23.574$ & $-23.495$ & $-23.587$ & $-23.553$\\ 

\hline
\end{tabular}
\caption{Central luminosity-$V_{max}$ relation from SHAM for all the models.}
\label{table:L-Vmax_central}
\end{table*}

\begin{table*}
\centering
\begin{tabular}{c c c c c c c}
\hline
\hline
$V_{max}$ & $M_{r}-5\log_{10}h$(GR) & $M_{r}-5\log_{10}h$(F4) &$M_{r}-5\log_{10}h$(F5) &$M_{r}-5\log_{10}h$(F6)& $M_{r}-5\log_{10}h$(N1) & $M_{r}-5\log_{10}h$(N5) \\
\hline
\hline
$302.755$ & $-21.5822$ & $-21.56465$ & $-21.49528$ & $-21.56465$ & $-21.54608$ & $-21.57182$\\
$319.467$ & $-21.60214$ & $-21.5881$ & $-21.51072$ & $-21.5881$ & $-21.5675$ & $-21.59634$\\
$337.102$ & $-21.63912$ & $-21.63035$ & $-21.54168$ & $-21.63035$ & $-21.60684$ & $-21.64038$\\
$355.707$ & $-21.692$ & $-21.68695$ & $-21.58824$ & $-21.68695$ & $-21.66222$ & $-21.70016$\\
$375.341$ & $-21.75454$ & $-21.7507$ & $-21.64654$ & $-21.7507$ & $-21.72714$ & $-21.76848$\\
$396.061$ & $-21.8218$ & $-21.81585$ & $-21.71232$ & $-21.81585$ & $-21.79584$ & $-21.83958$\\
$417.919$ & $-21.88932$ & $-21.87905$ & $-21.78154$ & $-21.87905$ & $-21.86458$ & $-21.90962$\\
$440.987$ & $-21.95426$ & $-21.94015$ & $-21.85288$ & $-21.94015$ & $-21.93154$ & $-21.97776$\\
$465.331$ & $-22.01804$ & $-22.0007$ & $-21.92504$ & $-22.0007$ & $-21.99714$ & $-22.04538$\\
$491.012$ & $-22.08202$ & $-22.06215$ & $-21.99718$ & $-22.06215$ & $-22.06242$ & $-22.11334$\\
$518.115$ & $-22.14584$ & $-22.12445$ & $-22.06986$ & $-22.12445$ & $-22.1279$ & $-22.18184$\\
$546.716$ & $-22.21048$ & $-22.18785$ & $-22.14178$ & $-22.18785$ & $-22.19358$ & $-22.2503$\\
$576.888$ & $-22.2767$ & $-22.25185$ & $-22.21364$ & $-22.25185$ & $-22.2602$ & $-22.31898$\\
$608.732$ & $-22.34306$ & $-22.3167$ & $-22.28684$ & $-22.3167$ & $-22.3284$ & $-22.38802$\\
$642.335$ & $-22.40992$ & $-22.3836$ & $-22.36016$ & $-22.3836$ & $-22.39646$ & $-22.45708$\\
$677.785$ & $-22.47814$ & $-22.45205$ & $-22.4347$ & $-22.45205$ & $-22.46474$ & $-22.52822$\\
$715.197$ & $-22.54396$ & $-22.522$ & $-22.50862$ & $-22.522$ & $-22.53216$ & $-22.60016$\\
$754.678$ & $-22.60604$ & $-22.5935$ & $-22.57768$ & $-22.5935$ & $-22.59602$ & $-22.67036$\\
$796.328$ & $-22.66382$ & $-22.6626$ & $-22.64194$ & $-22.6626$ & $-22.65844$ & $-22.73892$\\
$840.284$ & $-22.71466$ & $-22.72995$ & $-22.69932$ & $-22.72995$ & $-22.72122$ & $-22.8023$\\
$886.669$ & $-22.76278$ & $-22.7961$ & $-22.75238$ & $-22.7961$ & $-22.78622$ & $-22.8619$\\
$935.604$ & $-22.81816$ & $-22.8612$ & $-22.81002$ & $-22.8612$ & $-22.8574$ & $-22.92396$\\
$987.247$ & $-22.88942$ & $-22.9299$ & $-22.88042$ & $-22.9299$ & $-22.93414$ & $-22.9915$\\
$1041.746$ & $-22.97908$ & $-23.00385$ & $-22.96578$ & $-23.00385$ & $-23.01592$ & $-23.06668$\\
$1099.238$ & $-23.0853$ & $-23.08025$ & $-23.06422$ & $-23.08025$ & $-23.09606$ & $-23.14878$\\
$1159.914$ & $-23.18358$ & $-23.15195$ & $-23.15592$ & $-23.15195$ & $-23.16382$ & $-23.2234$\\
$1223.945$ & $-23.25788$ & $-23.2076$ & $-23.2252$ & $-23.2076$ & $-23.21366$ & $-23.28066$\\
$1291.493$ & $-23.29234$ & $-23.23805$ & $-23.25996$ & $-23.23805$ & $-23.2401$ & $-23.3115$\\
$1362.781$ & $-23.27244$ & $-23.23515$ & $-23.24584$ & $-23.23515$ & $-23.23878$ & $-23.30766$\\
\hline
\end{tabular}
\caption{Satellites luminosity-$V_{\rm max}$ relation from SHAM for all the models.}
\label{table:L-Vmax_satellite}
\end{table*}

\end{document}